\documentclass[acmsmall]{acmart}

\usepackage{framed}
\usepackage{multirow}
\usepackage{booktabs}
\usepackage{ifthen}
\usepackage{color}
\usepackage{url}
\usepackage{amsmath}
\usepackage{xspace}
\usepackage[T1]{fontenc}
\usepackage{enumerate}
\usepackage[linesnumbered,ruled,vlined]{algorithm2e}
\usepackage{array,multirow,graphicx}
\usepackage{float}
\usepackage{balance}
\usepackage{tikz}
\usepackage{calc}
\usepackage{subfigure}
\usepackage{listings,amsfonts}
\usepackage{amsmath,xcolor,pifont}
\usepackage{url}
\usepackage{xspace}
\usepackage{hyperref,endnotes}
\usepackage{xcolor}
\usepackage{array}
\usepackage{multirow,makecell}
\usepackage[misc]{ifsym}

\begin{document}

\setcopyright{acmcopyright}
\acmJournal{POMACS}
\acmYear{2021} \acmVolume{5} \acmNumber{3} \acmArticle{39} \acmMonth{12} \acmPrice{15.00}\acmDOI{10.1145/3491051}

\received{August 2021}
\received[revised]{October 2021}
\received[accepted]{November 2021}

\title{Trade or Trick? Detecting and Characterizing Scam Tokens on Uniswap Decentralized Exchange}

\author{Pengcheng Xia}
\affiliation{%
  \institution{Beijing University of Posts and Telecommunications}
  \city{Beijing}
  \country{China}
}

\author{Haoyu Wang}
\authornote{Corresponding Author: Haoyu Wang (haoyuwang@bupt.edu.cn).}
\affiliation{%
  \institution{Beijing University of Posts and Telecommunications}
  \city{Beijing}
  \country{China}
}
\email{haoyuwang@bupt.edu.cn}

\author{Bingyu Gao}
\affiliation{%
  \institution{Beijing University of Posts and Telecommunications}
  \city{Beijing}
  \country{China}
}

\author{Weihang Su}
\affiliation{%
  \institution{Beijing University of Posts and Telecommunications}
  \city{Beijing}
  \country{China}
}

\author{Zhou Yu}
\affiliation{%
  \institution{Beijing University of Posts and Telecommunications}
  \city{Beijing}
  \country{China}
}

\author{Xiapu Luo}
\affiliation{%
  \institution{The Hong Kong Polytechnic University}
  \city{HongKong}
  \country{China}
}

\author{Chao Zhang}
\affiliation{%
  \institution{Tsinghua University}
  \city{Beijing}
  \country{China}
}

\author{Xusheng Xiao}
\affiliation{%
  \institution{Case Western Reserve University}
  \city{Cleveland}
  \country{United States}
}

\author{Guoai Xu}
\affiliation{%
  \institution{Beijing University of Posts and Telecommunications}
  \city{Beijing}
  \country{China}
}

\renewcommand{\shortauthors}{Pengcheng Xia et al.}

\begin{abstract}

The prosperity of the cryptocurrency ecosystem drives the need for digital asset trading platforms. Beyond centralized exchanges (CEXs), decentralized exchanges (DEXs) are introduced to allow users to trade cryptocurrency without transferring the custody of their digital assets to the middlemen, thus eliminating the security and privacy issues of traditional CEX. Uniswap, as the most prominent cryptocurrency DEX, is continuing to attract scammers, with fraudulent cryptocurrencies flooding in the ecosystem. 
In this paper, we take the first step to detect and characterize scam tokens on Uniswap. 
We first collect all the transactions related to Uniswap V2 exchange and investigate the landscape of cryptocurrency trading on Uniswap from different perspectives. Then, we propose an accurate approach for flagging scam tokens on Uniswap based on a guilt-by-association heuristic and a machine-learning powered technique. We have identified over 10K scam tokens listed on Uniswap, which suggests that roughly 50\% of the tokens listed on Uniswap are scam tokens.
All the scam tokens and liquidity pools are created specialized for the ``rug pull'' scams, and some scam tokens have embedded tricks and backdoors in the smart contracts.
We further observe that thousands of collusion addresses help carry out the scams in league with the scam token/pool creators.
The scammers have gained a profit of at least \$16 million from 39,762 potential victims. 
Our observations in this paper suggest the urgency to identify and stop scams in the decentralized finance ecosystem, and our approach can act as a whistleblower that identifies scam tokens at their early stages.

\end{abstract}

\begin{CCSXML}
<ccs2012>
<concept>
<concept_id>10002978.10002997</concept_id>
<concept_desc>Security and privacy~Intrusion/anomaly detection and malware mitigation</concept_desc>
<concept_significance>500</concept_significance>
</concept>
<concept>
<concept_id>10002978.10003022.10003026</concept_id>
<concept_desc>Security and privacy~Web application security</concept_desc>
<concept_significance>500</concept_significance>
</concept>
<concept>
<concept_id>10002951.10003260.10003277</concept_id>
<concept_desc>Information systems~Web mining</concept_desc>
<concept_significance>500</concept_significance>
</concept>
</ccs2012>
\end{CCSXML}

\ccsdesc[500]{Security and privacy~Intrusion/anomaly detection and malware mitigation}
\ccsdesc[500]{Security and privacy~Web application security}
\ccsdesc[500]{Information systems~Web mining}

\keywords{Uniswap; scam cryptocurrency; exchange; blockchain}

\maketitle

\section{Introduction}

Cryptocurrencies have seen significant growth in recent years due to the rapid development of blockchain technologies and the digital economic system. By the end of July 2021, the global cryptocurrency market capitalization reaches over \$ 1.5 trillion~\cite{coinmarketcap}. Thousands of cryptocurrencies and decentralized applications (DApps) are emerging in the ecosystem. 

The prosperity of the cryptocurrency ecosystem drives the need for digital asset trading platforms.
Thus, hundreds of cryptocurrency exchanges are emerging to facilitate the trading of digital assets. Cryptocurrency exchanges can be categorized into two types: centralized exchange (CEX) and decentralized exchange (DEX). 
CEX, as the traditional trading mechanism, requires a central entity as the intermediary to complete cryptocurrency trading between its users. Therefore, the trustworthiness of the middlemen plays a vital role in this trading mechanism, as all the user activities and digital assets are under the control of the central operators\footnote{Most CEXs have adopted Know Your Customer (KYC) verification to prevent money laundering and other financial crimes.}. Security and privacy issues of CEXs are reported from time to time~\cite{binancehack,buyucoinleak,kucoinhack}.
To facilitate free trading and eliminate the potential security and privacy issues, 
DEX is introduced to allow users to trade their cryptocurrencies without transferring the custody of their cryptocurrencies to the middlemen, thereby mitigating the security issues of CEX and providing better privacy by eliminating KYC verification.

Uniswap is one of the most prominent cryptocurrency DEXs built atop the Ethereum blockchain~\cite{uniswap}. 
Unlike most exchanges, which match buyers and sellers to determine prices and execute trades, Uniswap adopts the automated market maker (AMM) model~\cite{amm}. This model involves smart contracts creating liquidity pools of cryptocurrencies that are automatically traded based on pre-set algorithms.
As a DEX, anyone can use the pools to swap between cryptocurrencies for a small fee. In addition, users can also be liquidity providers by depositing cryptocurrencies into the liquidity pools and earn said swap fees as incentives. By the time of this study, Uniswap has amassed total market liquidity of over \$ 1.6 Billion, with over \$ 200 Million trading volume per day~\cite{uniswapinfo}. 

\textit{Where there is money, there are those who follow it}. The growing popularity of Uniswap is continuing to attract scammers. Uniswap does not maintain any rules or criteria for cryptocurrency listing, meaning that anybody can list a token on the exchange. Thus, scammers take the opportunity to list scam cryptocurrencies to trick unsuspecting users. It was reported that some scam cryptocurrencies impersonate token sales for popular cryptocurrency projects~\cite{scam_news,scam1,scam2}. For example, on August 19th 2020, the upcoming DeFi lending protocol \texttt{Teller Finance} tweeted that a fake Teller token and a scam Uniswap pool had been created, and many users were cheated.

\textit{Despite this, to the best of our knowledge, no previous studies have systematically characterized or measured scam tokens on Uniswap}. We are unaware of to what extent scam tokens exist on the Uniswap exchange, and how much impact they introduced to the overall ecosystem. 

\textbf{This Work.}
In this paper, we take the first step to detect and characterize scam tokens on Uniswap. We first collect all the transactions related to the Uniswap exchange, and investigate the landscape of Uniswap from different perspectives (see \textbf{Section~\ref{sec:general}}). Then, we propose a hybrid approach for flagging scam tokens and scam liquidity pools on Uniswap accurately (see \textbf{Section~\ref{sec:detection}}). 
We manually label a scam token benchmark dataset, and identify features that can be used to distinguish them. Our detection approach is powered by a guilt-by-association based expanding method, and a machine-learning based detection and verification technique. We have identified over 10K scam tokens and pools in total (which is a lower-bound), meaning that roughly 50\% of tokens listed on Uniswap are scam tokens. 
At last, we demystify these scam tokens from various perspectives, including the scam behaviors, the scammers, and the impact (see \textbf{Section~\ref{sec:analysis}}). 
Beyond the scam tokens and their creators, we further identify over 40K collusion addresses controlled by the scammers, which are used to facilitate the success of the scams.
We show that, the scammers at least profit \$16 million from roughly 40K potential victims on Uniswap.

We make the following main research contributions in this paper:
\begin{itemize}
    \item \textit{We are the first to propose a reliable approach for identifying scam tokens and their associated liquidity pools on Uniswap}. We first make effort to contribute by far the largest scam token benchmark dataset, and then we propose a guilt-by-association based expansion method and a machine-learning based classifier to identify the most reliable scam tokens. 
    
    \item \textit{We identify over 10K scam tokens and scam liquidity pools, revealing the shocking fact that Uniswap is flooded with scams}. We believe the scams are prevalent on other DEXs and DeFi platforms, due to the inherent loose regulation of the decentralized ecosystem.
    
    \item \textit{We systematically characterize the behaviors, the working mechanism, and the financial impacts of Uniswap scams.}
    We observe that scammers usually employ multiple addresses to carry out a scam, and thus we design a method for detecting the collusion addresses to gain a deep understanding of the scams. We have identified $70,331$ scam addresses in total, including $41,118$ collusion addresses. The scammers have gained at least \$16 millions.
\end{itemize}

This is the first in-depth study of Uniswap scams at scale, longitudinally and across various dimensions. 
Our results motivate the need for more research efforts to illuminate the widely unexplored scams in the decentralized finance ecosystem. 
We advocate stakeholders in the ecosystem work together to eliminate the impact caused by scam tokens. Basically, a cryptocurrency reputation system is needed, and our approach can be adopted to identify scam tokens at their early stages. Blockchain services like wallets and exchanges should provide useful front-end to warn users when they try to engage with the high-risk tokens. Further, investors and Defi project teams should be aware of the scam tokens and rely on trusted sources to make decisions.

\section{Background}

\subsection{Blockchain and Ethereum}

Blockchain is a shared, immutable, and distributed ledger that facilitates the process of recording transactions and tracking assets in a P2P network~\cite{blockchain}. It is resistant to data modification due to the cryptographic design. By this design, each transaction in the block is verified by the confirmation of most participants in the system. Bitcoin network is the first blockchain-based decentralized system, which demonstrated the feasibility to construct a decentralized value-transfer system that can be shared across the world and virtually free to use. 

Following the growth of cryptocurrencies, developers started to explore the feasibility of decentralized applications (DApps)~\cite{dapp}. This leads to the development of Ethereum~\cite{ethereum}, an open-source decentralized blockchain platform featuring smart contract functionality. \textit{Ether} (ETH) is the official cryptocurrency on Ethereum, which is mined by Ethereum miners as a reward for computations. ETH is the second largest cryptocurrency based on the market cap~\cite{coinmarketcap}.

\subsection{Ethereum Accounts and Transactions}
\subsubsection{Accounts} 
In Ethereum, an account is a basic unit to identify an entity. An account is identified by a fixed-length hash-like address. Accounts can be user-controlled or deployed as smart contracts.
For the accounts that are controlled by users, i.e., by anyone with private keys, they are called \textit{external owned accounts} (EOAs). The accounts controlled by code are called \textit{contract owned accounts} (COAs). Both kinds of accounts have the ability to send, receive, hold ETH and tokens, or interact with deployed smart contracts. The key difference is that, only an EOA can initiate transactions while a COA can only send transactions in response to receiving transactions.

\subsubsection{Transactions on Ethereum} 
A transaction refers to an action initiated by an EOA and it is the way that users interact with the Ethereum network. A transaction is used to modify or update the state stored in the Ethereum network and it requires a fee and must be mined to become valid.
A transaction can include binary data (called the ``payload'') and Ether. If a transaction is sent from an EOA to another EOA, the transaction is called ``external transaction'', which will be included in the blockchain and can be obtained by parsing the blocks. The other type of transaction, initiated by executing a smart contract, is called ``internal transaction''. Internal transactions are usually triggered by external transactions and are not stored in the blockchain directly. When smart contracts are involved in a transaction, multiple events that log the running status of contracts could be emitted for developers and DApps to track the behavior of these contracts.

\subsection{Smart Contract and ERC-20 Token}

\subsubsection{Smart Contract}
A smart contract is a computer program or a transaction protocol that can execute automatically with the terms of the agreement written in the contract code. The contract code controls the execution, and the corresponding transactions can be tracked but cannot be reversed. 
Ethereum implements a Turing-complete language on its blockchain, and now it is the largest blockchain platform that supports smart contracts with millions of deployed smart contracts. 

\subsubsection{ERC-20 Token}
In contrast to digital coins like Bitcoin and Ether, which are native to their own blockchain, ``tokens'' require existing blockchain platforms. 
On the Ethereum platform, there are over 400K tokens by the time of this study, and most of them are smart contracts following the ERC-20 standard\footnote{ERCs stands for Ethereum Requests for Comments, which are technical documents used by smart contract developers.}, which specifies a list of rules and interfaces that tokens should follow.
Some of these rules include the total supply of the tokens, how the tokens are transferred and how the transactions are approved, etc. 
Note that, Ethereum does not enforce any restrictions on the names and symbols of tokens, which may open doors for scammers to abuse the ERC-20 tokens. We will show that, due to the less regulation of Uniswap and Ethereum, scam ERC-20 tokens are prevalent in the ecosystem (see Section~\ref{sec:detection} and Section~\ref{sec:analysis}). To remove ambiguity, in this paper, we will describe a token in the form of ``name (symbol)'' with a footnote of token address.

\subsection{DEX, AMM, and Uniswap}

\subsubsection{Decentralized Exchange}
Due to the open-source and decentralized nature of cryptocurrencies, it is demanded that the exchange of cryptocurrencies should have no central authorities involved, and thus decentralized exchanges (DEXs) are born.
A blockchain-based DEX does not store user funds and personal data on centralized servers, but instead matches buyers and sellers of digital assets through smart contracts.
DEXs are an important part of the burgeoning DeFi ecosystem.

There are multiple kinds of DEXs. The first generation is order-based P2P exchange, which uses order books. These order books compile a record of all open buy and sell orders for a particular asset. For example, dYdX~\cite{dYdX}, IDEX~\cite{IDEX}, and EtherDelta~\cite{EtherDelta} fall to this category. The second generation is a liquidity pool based exchange that completes trades through automated market makers (AMMs). The representative ones are Uniswap~\cite{uniswap}, Bancor~\cite{Bancor}, and Balancer~\cite{balancer}.

\subsubsection{Automated Market Makers.} 
Automated market makers (AMMs) allow digital assets to be traded without permission and automatically by using liquidity pools instead of a traditional market of buyers and sellers. On the AMM markets, users trade against a pool of tokens, i.e., a liquidity pool. Users can supply tokens into the liquidity pools and the price of tokens in the pool is determined by a mathematical formula. Liquidity providers normally earn a fee for providing tokens to the pool, and the fee is paid by the traders who interact with the pool.

\subsubsection{Uniswap.}
Uniswap is a leading DEX built atop Ethereum designed to facilitate automated exchange transactions between ETH and ERC-20 tokens, providing liquidity automatically on Ethereum. 
Uniswap is the largest decentralized exchange and the fourth-largest cryptocurrency exchange overall by daily trading volume by the time of this study.

Uniswap V1, the first version of the protocol, was created in November 2018 by Hayden Adams, and it supports all the ETH to ERC-20 liquidity pools and enables swaps between ERC-20 tokens via ETH. In May 2020, Uniswap launched its V2 version with many new features and optimizations. For example, it uses WETH (Wrapped ETH, an ERC-20 token that represents ETH 1:1) instead of the native ETH in its core contracts and enables direct ERC-20 to ERC-20 swaps, thus halving the fees when performing such transactions. Also, it enables non-standard ERC-20 tokens such as USDT and BNB, which opened up the potential market. Furthermore, the Uniswap V2 introduces ``flash swap'', which allows users to borrow tokens from a Uniswap pool, perform some activities with external services and pay back these tokens, like flash loans. These changes set the stage for exponential growth in AMM adoption. In May 2021, Uniswap V3 was launched, which provides new features like concentrated liquidity and multiple fee tiers, making the protocol more flexible and efficient. Since Uniswap V3 was just launched and these three versions of Uniswap operate independently, Uniswap V2 remains the most popular one by the time of this study, 
with a large number of tokens locked in it. Although our study in this paper is focused on Uniswap V2, the observations and implications are applicable to other versions of Uniswap and other DEXs.

\subsection{Interacting with Uniswap}
\begin{figure}[t]
    \centering
    \includegraphics[width= 0.6\linewidth]{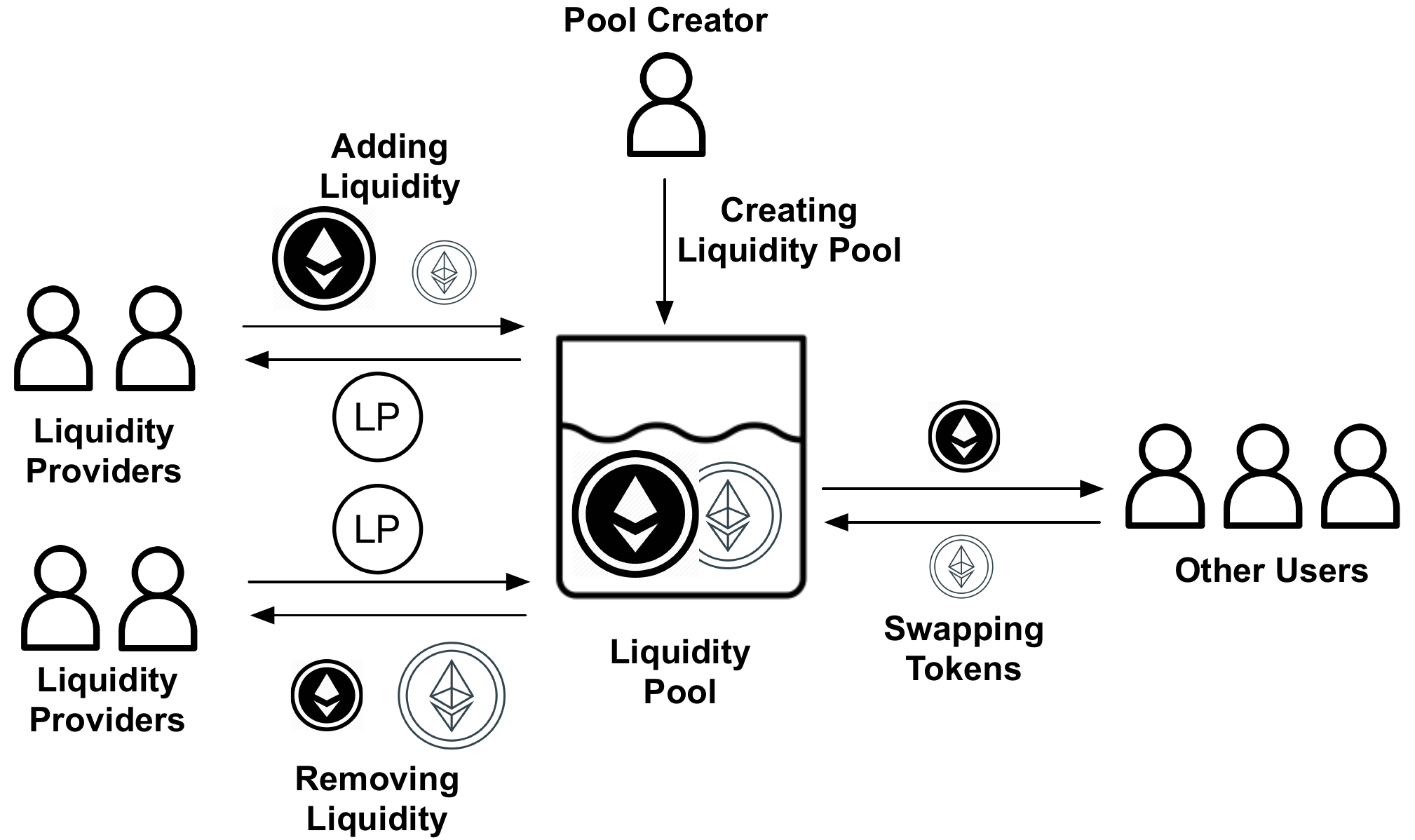}
    \caption{Interacting with Uniswap V2 and the major participants.}
    \label{fig:majorpar}
\end{figure}

Users can interact with Uniswap through three kinds of operations, i.e., creating a liquidity pool, adding/removing liquidity, and swapping tokens. The general process is shown in Figure~\ref{fig:majorpar}. 

\subsubsection{Creating the liquidity pools.} In Uniswap, users trade against liquidity pools. A liquidity pool is a trading venue for a pair of ERC-20 tokens. One can create a liquidity pool that does not exist by interacting with Uniswap V2 contracts. 

\subsubsection{Adding/removing liquidity.} After the pool is created, users can add liquidity by depositing the pair of two tokens in the pool. The users who add liquidity to the pool are called \texttt{liquidity providers} (LPs for short) and they will receive \texttt{liquidity provider tokens} (LP tokens for short). A ``mint'' event will be emitted when liquidity is added. Whenever a trade occurs, a 0.3\% fee is charged to the transaction sender. This fee is distributed pro-rata to all LPs in the pool upon completion of the trade, which stimulates people to provide liquidity.

In order for the pool to begin facilitating trades, someone must seed it with an initial deposit of each token after a pool's creation. Thus, the pool creator will usually add the first liquidity when creating the pool. According to the documents of Uniswap~\cite{uniswapdoc}, if the first LP supplies $x$ A tokens and $y$ B tokens, he will receive $l = \sqrt{x*y}$ LP tokens and the total supply of the pool token is $l$. 
If there are already $x$ A tokens and $y$ B tokens in the pool, the new LP could supply $x^{'}$ A tokens and $y^{'}$ B tokens based on the current ratio and he will receive $l^{'} =l * \frac{x^{'}}{x}$ LP tokens and the total supply of LP token changes to $l + l^{'}$.

The LPs could also remove liquidity from the pool by burning their LP tokens. After removing the liquidity, they can receive the pair of tokens based on the LP tokens they burn and the current token supply in the pool. A ``burn'' event will be emitted when LP tokens are burned.
For example, if there are $x$ A tokens and $y$ B tokens in the pool and the total supply of LP token is $l$, when an LP burns $l^{'}$ LP token, he will receive $x^{'}$ A tokens and $y^{'}$ B tokens where $\frac{x^{'}}{x} = \frac{y^{'}}{y} = \frac{l^{'}}{l}$ and the total supply of LP tokens will be $l-l^{'}$.

\subsubsection{Swapping tokens.}
When a user wants to trade a pair of tokens in a pool, the user will first send tokens to the pool. Then the pool will calculate the exchange rate and send the target tokens. The exchange rate is determined by the ``constant product'' formula $k = x * y$, where $k$ is a constant and $x,y$ are the reserve balance of two tokens in the pool. In a swap transaction, the LP token will not change and a ``swap'' event will be emitted. For example, if the pool has $x$ A tokens and $y$ B tokens and the user sends $x^{'}$ A tokens for B tokens. The swap will follow Eqn. (\ref{eqn:swapequation}):
\begin{equation}
\label{eqn:swapequation}
x*y=(x+x^{'}*0.997) * (y-y^{'}), 
\end{equation}
where $0.997$ implies the 0.3\% of fee set by Uniswap and $y^{'}$ will be the quantity of B tokens the user gets. Due to this formula, one token's price in the pool will rise when people are swapping the other token for this one.

\section{General Overview of the Uniswap Exchange}
\label{sec:general}

\subsection{Dataset Collection}

\subsubsection{Collection Method}
\label{sec:dataset_collection}
We utilize \texttt{The Graph}~\cite{thegraph}, a sandbox for querying data and endpoints for blockchain developers, to collect transaction events (i.e., mint, swap, and burn events) related to Uniswap. It provides a snapshot of the current state of Uniswap and also tracks the historical data.
Note that an Ethereum transaction can emit multiple events including Uniswap-related events and other events, but \texttt{The Graph} only records all data related to Uniswap events. 
Moreover, some important information is not recorded in the log. For example, when a user interacts with the Uniswap router contract for trading some tokens to Ethers (not WETH), the router contract will transfer the tokens on behalf of the user, exchange the WETH to Ethers, and transfer the Ethers to the user. Thus, the log will record the router contract as the swap event receiver instead of the user. 
As we need to analyze the detailed token transfer flow of transactions related to Uniswap for characterizing the scam token activities in Section~\ref{sec:analysis}, we further fetch the whole transaction information related to events on Uniswap, e.g., the amount of ETH transferred, input data, internal transactions, and all event logs.

\begin{table}[h]
\small
\caption{Dataset Overview.}
%\resizebox{0.45\linewidth}{!}{
\begin{tabular}{@{}cr||cr@{}}
\toprule
Data Type & \# of Entities & Event Type & \# of Events \\ \midrule
Token       & 21,778     & Mint & 804,077    \\
Pair (pool)        & 25,131     & Burn & 415,919    \\
Events & 20,306,762 & Swap & 19,086,766 \\ \bottomrule
\end{tabular}
%}
\label{tab:data}
\end{table}

\subsubsection{Dataset Overview}
We have synchronized all the tokens and events from May 5th 21:00 UTC to December 6th 18:00 UTC, 2020.
Table~\ref{tab:data} shows an overview of our dataset.
Since the first transaction which created \texttt{USD Coin (USDC)-Wrapped Ether (WETH)} liquidity pool happened on May 5th, 2020, there are over 20 million transaction events on Uniswap V2 by the time of this study. There are $21,778$ kinds of tokens and $25,131$ liquidity pools in total.

\begin{figure*}[h]\centering 
\subfigure[Token listing and Pool Creation.]{ 
\begin{minipage}{0.31\textwidth}\centering
\includegraphics[width = 0.99\textwidth]{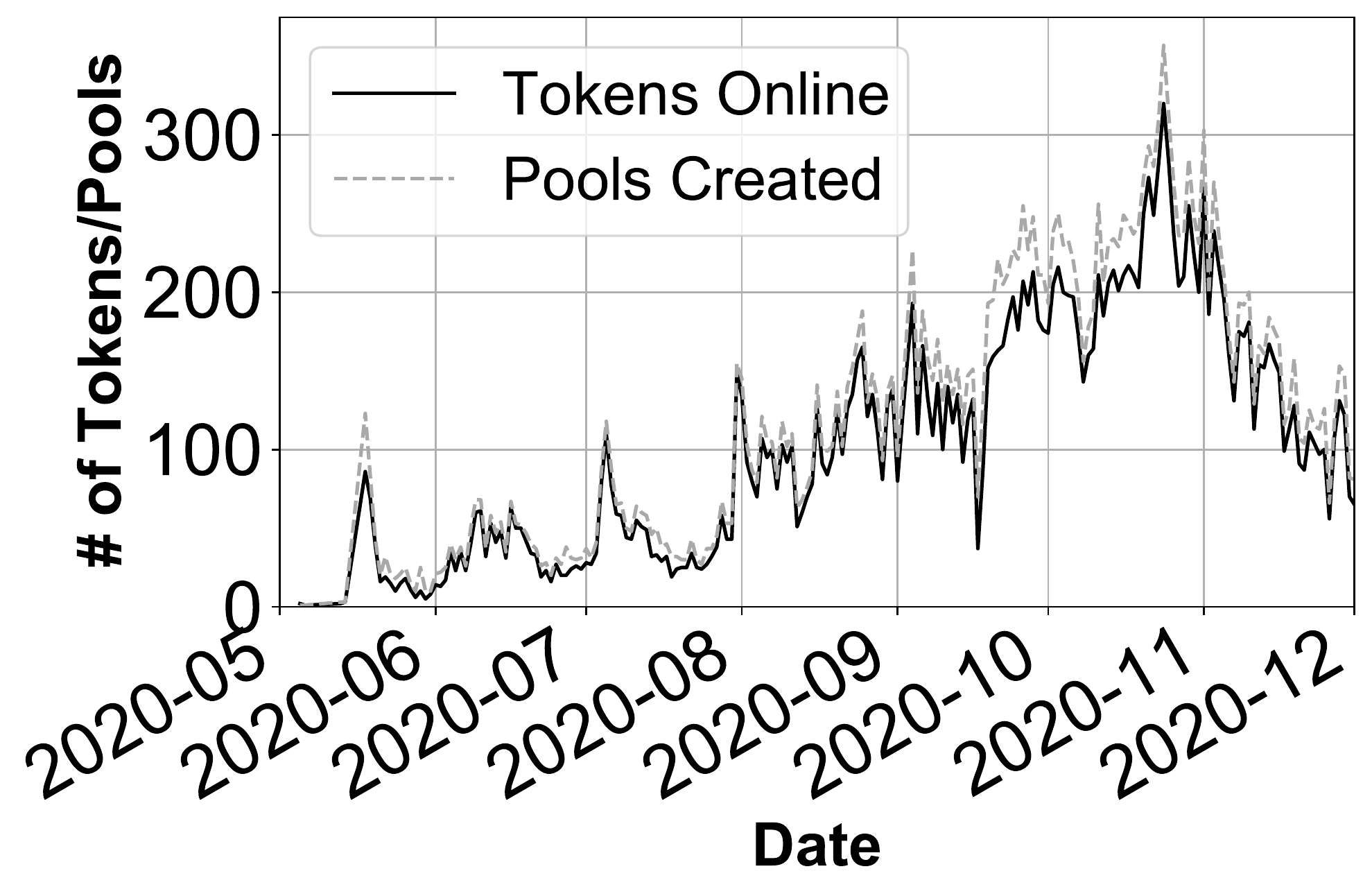} 
\end{minipage}}
\hspace{0.03in}
\subfigure[Transaction Events]{
\begin{minipage}{0.31\textwidth}
\centering 
\includegraphics[width = 0.99\textwidth]{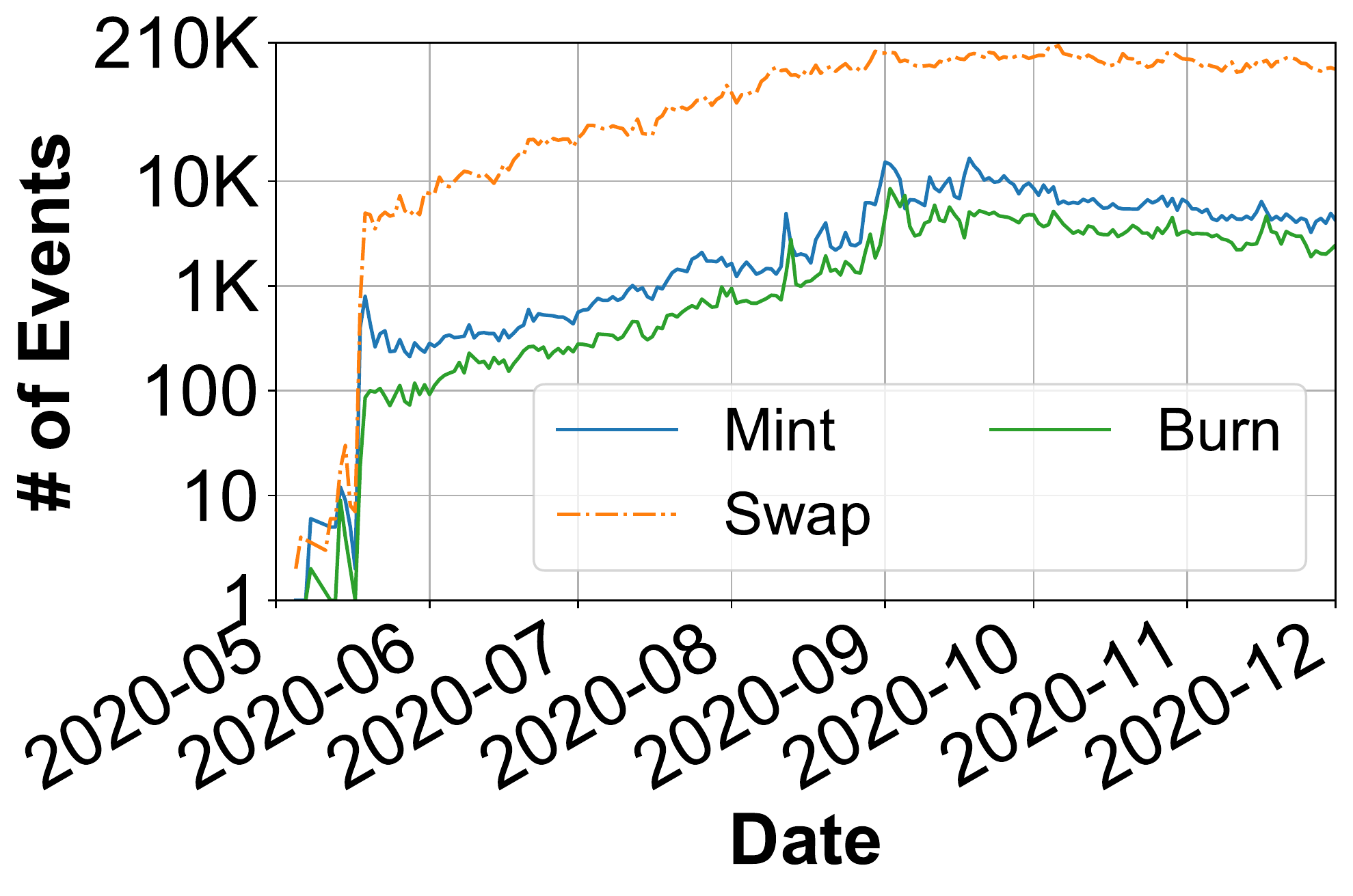} 
\end{minipage}
}
\hspace{0.03in}
\subfigure[Transaction Volume and Liquidity]{ 
\begin{minipage}{0.31\textwidth}
\centering 
\includegraphics[width = 0.99\textwidth]{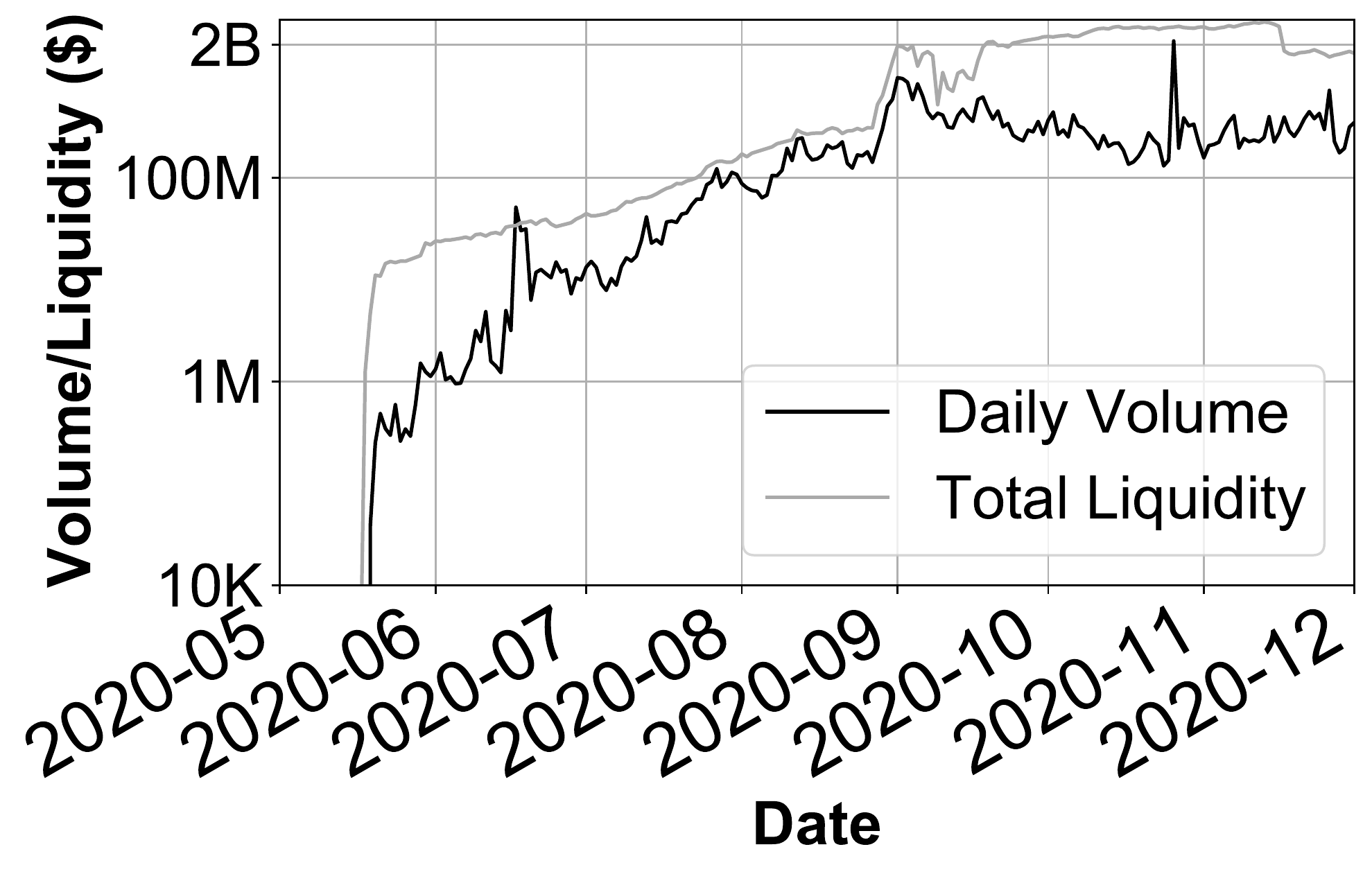} 
\end{minipage}
}
\caption{The general trend of Uniswap V2 from May 2020 to December 2020 (daily).}
\label{fig:evolution}
\end{figure*}

\subsection{The Rising of Uniswap}
\label{sec:uniswap_evo}

Figure~\ref{fig:evolution} (a) shows the daily \textit{token listing} (i.e., appear on Uniswap, not token creation) and \textit{liquidity pool creation} on Uniswap. After three months of the launch of Uniswap V2, there are roughly over 100 tokens and liquidity pools listed on Uniswap daily. It witnessed the spike of tokens and pairs (pools) on Uniswap in October 2020, when $6,677$ tokens were added and $7,546$ pools were created in this month. 
Figure~\ref{fig:evolution} (b) shows the number of transaction events relevant to Uniswap. Since its launch, Uniswap attracted great attention quickly. It remains roughly 200K transaction events daily by the time of this study. For example, on October 6th, there were $207,338$ transaction events on Uniswap with $2,288$ pairs traded. 
It is not surprising to see swap transactions, as a major function of Uniswap, account for 94\% of the total events. 
Figure~\ref{fig:evolution} (c) shows the daily volume and total liquidity. After August 7th, the daily volume exceeded \$100 million. The liquidity of Uniswap reached \$3.4 billion on 13th November and dropped due to the end of Uniswap (UNI) liquidity program~\cite{liquiditym}\footnote{In this program that ended on November 17th, users can earn UNI by adding liquidity to the 4 major liquidity pools.}. Nevertheless, by the time of this study, there were still over \$1.7 billion worth of tokens locked in Uniswap V2.

\subsection{The Liquidity Pools and Tokens}
\label{sec:poppair}

\subsubsection{Liquidity Pools.}

\begin{figure}[h]
\begin{minipage}[t]{0.31\linewidth}
    \includegraphics[width = 1\linewidth]{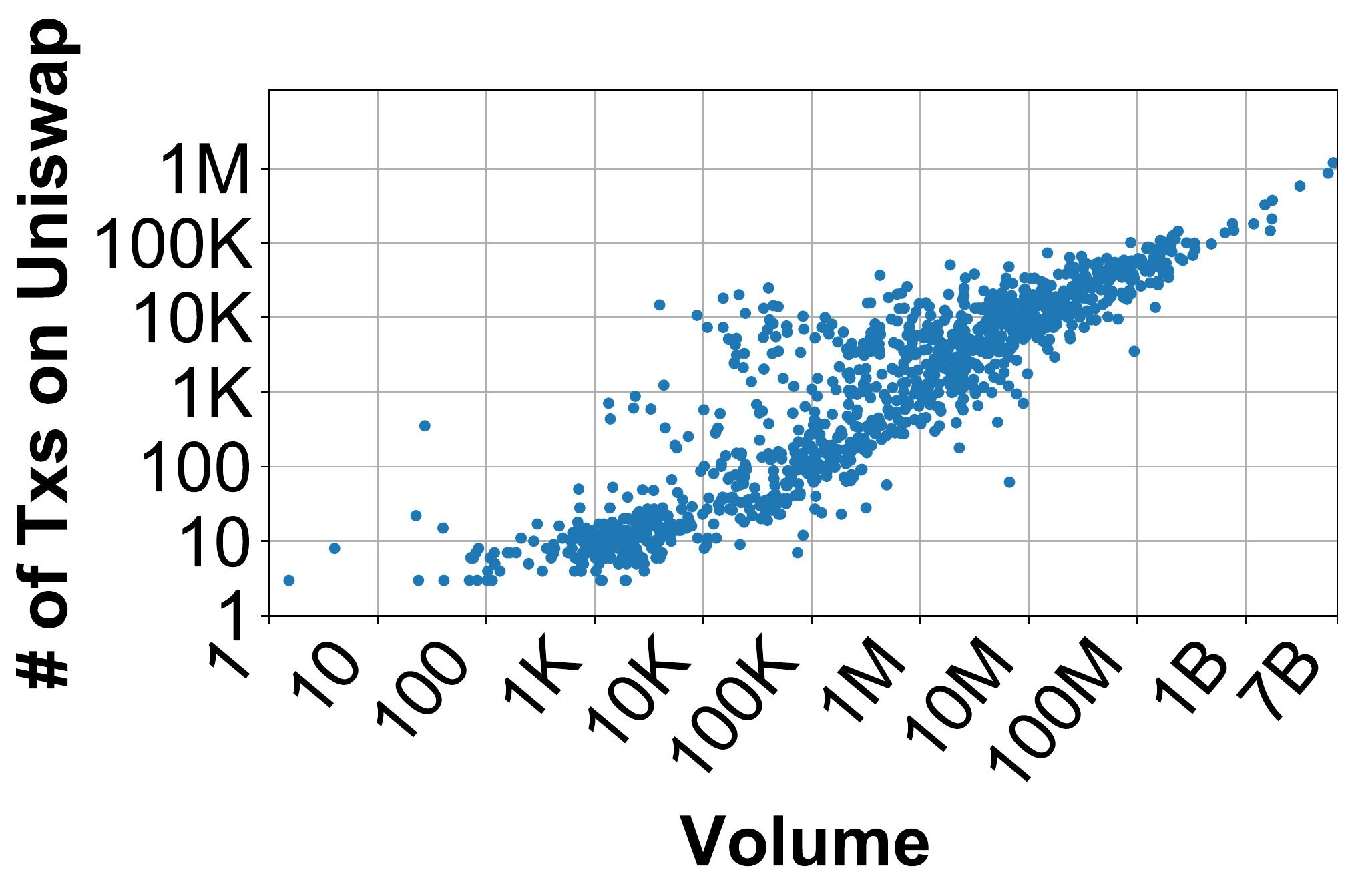}
    \caption{The event vs. total volume of liquidity pools on Uniswap.}
    \label{fig:tx}
\end{minipage}
\hspace{0.05in}
\begin{minipage}[t]{0.31\linewidth}
    \includegraphics[width = 1\linewidth]{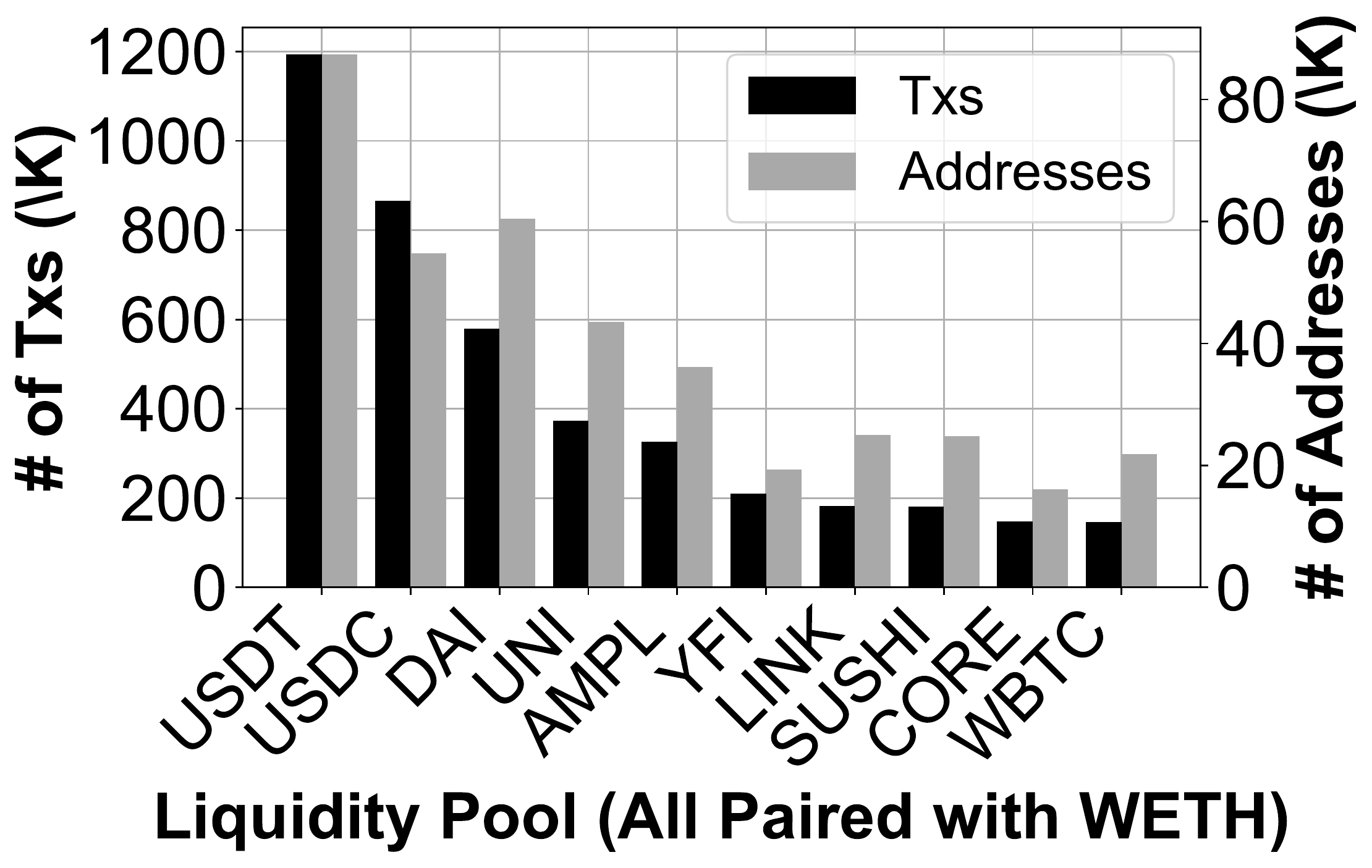}
    \caption{The top-10 popular pools on Uniswap.}
    \label{fig:poppair}
\end{minipage}
\hspace{0.05in}
\begin{minipage}[t]{0.31\linewidth}
    \includegraphics[width = 1\linewidth]{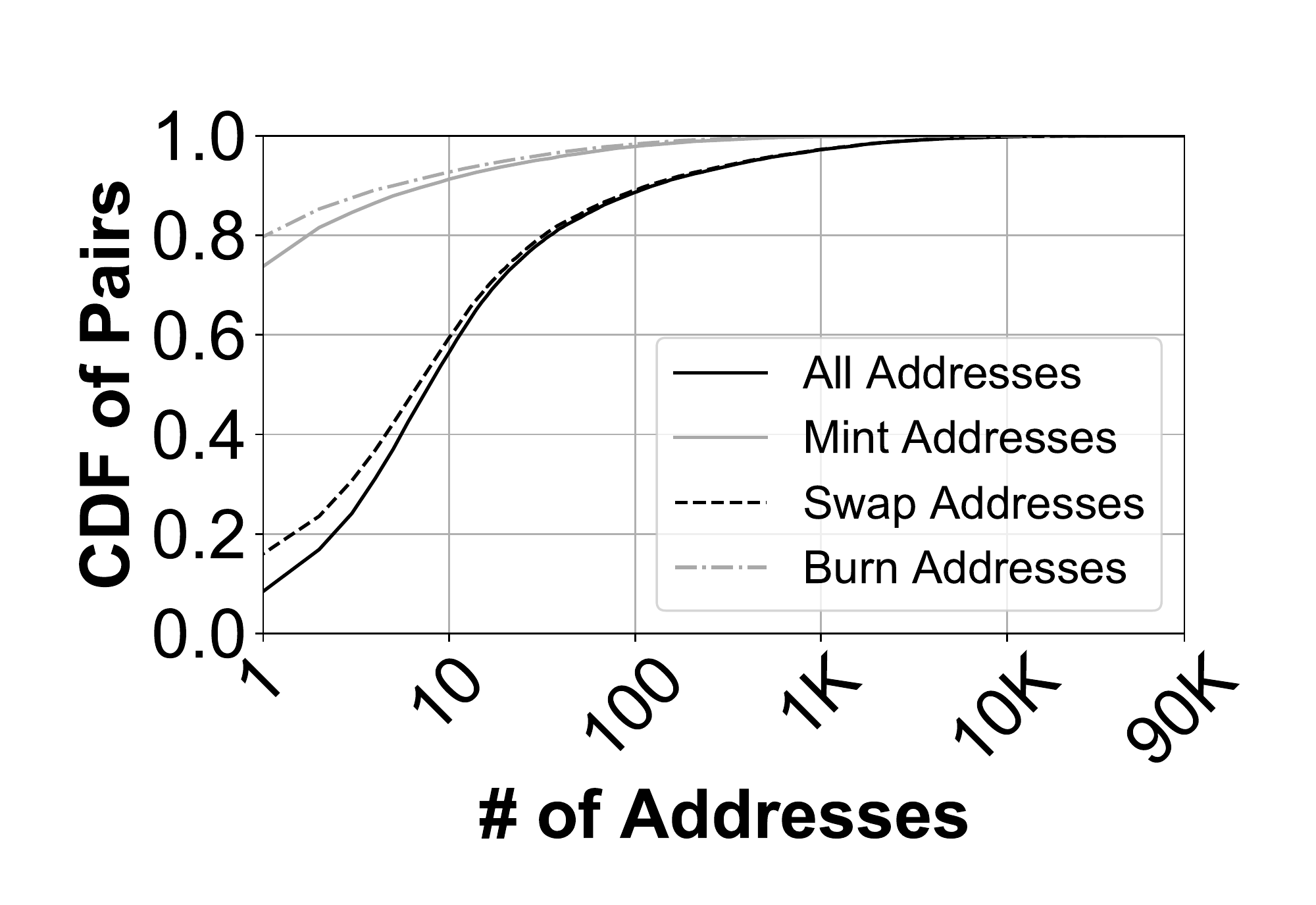}
    \caption{The involved addresses distribution of pairs on Uniswap.}
    \label{fig:pairaddr}
\end{minipage}
\end{figure}

We observe that over 90\% ($22,660$) of the liquidity pools have a value of less than 1 USD locked in Uniswap, which means that these pairs have low levels of liquidity on Uniswap or have low values. For example, although the pool \texttt{LiquidityBomberB (LBB)- LiquidityBomberA (LBA)}\footnote{LP token address:0xa0f198fc128b83c5f71cc61d105adf6c7d6fd88f} has a large number of tokens ($1\times10^{12}$ tokens), both of them have no value at all. The pair with the largest USD liquidity is \texttt{Wrapped Bitcoin (WBTC)- Wrapped Ether (WETH)}\footnote{LP token address:0xbb2b8038a1640196fbe3e38816f3e67cba72d940}, which locks over 195 million USD on Uniswap. 
Consequently, Uniswap only records the trade volume of pools with a certain level of liquidity, and over 95\% of the liquidity pools' volume is not recorded by Uniswap due to the lack of liquidity.
Figure~\ref{fig:tx} shows the distribution of transaction and trading volume for $1,128$ liquidity pools that have recorded trade volumes on Uniswap. They have a total trading volume of over \$41 billion. 
Obviously, it follows the typical Power-law distribution, i.e, the top 1\% of liquidity pools occupy over 65\% of the transaction events on Uniswap.
Figure~\ref{fig:poppair} shows the top-10 popular liquidity pools on Uniswap ranked by the transaction events. It can be seen that, stablecoins (e.g., Tether (USDT), USD Coin (USDC), Dai (DAI)) and Uniswap governance tokens (i.e., Uniswap (UNI) token), often have large popularity.

Figure~\ref{fig:pairaddr} shows the distribution of the involved address for each liquidity pool, which reflects the attention from investors. 
Over 70\% of the liquidity pools have been involved by less than 20 addresses and over 70\% of the liquidity pools have only 1 liquidity provider. As opposed to it, the 5 popular pairs all have more than 10K liquidity providers and the \texttt{Wrapped Ether (WETH)- Tether (USDT)} liquidity pool has more than 80K addresses involved.

\subsubsection{Tokens on Uniswap.}
From the perspective of ERC-20 tokens, the top 1\% of the tokens occupy over 80\% of the transactions and involve in over 96\% of the trading volume on Uniswap, which follows the Power-law effect as well. 
When considering popular tokens with the most number of liquidity pools, the stable coins also take the lead. Over 90\% (19,790) of the tokens only have one pair. In total, WETH is paired with over 20,924 tokens (83.2\% of all liquidity pools), followed by USDT (1,049), USDC (462), DAI (406), and UNI (253).

\subsection{Pool Creators and Investors} 
\label{sec:all_participant}

\subsubsection{Pool Creators}
All the $25,131$ liquidity pools analyzed were created by $17,053$ addresses. Among them, $3,046$ addresses had created at least 2 liquidity pools and 120 addresses created more than 10 pools. 
For example, the address \texttt{0x3bcfa9357ab84baec04313650d0eebb3fd51070d} created 91 liquidity pools and most of them are pairs of \texttt{WETH} and DeFi-related tokens such as ``Keep3r'', ``Wootrade Network'', ``Aegis.finance'', etc.
Due to the massive pool creation for various DeFi tokens, the address is suspicious to be a scammer and these pools are likely to be used in scams. We will further analyze these scams in Section~\ref{sec:analysis}.

\subsubsection{Investors}
\label{sec:investors}
In total, $548,609$ addresses have participated in the transactions collected in this study on Uniswap (i.e., the addresses that had mint, swap, burn transactions on Uniswap). Over 70\% ($387,885$) of the addresses only participate in swap transactions, and 5\% ($27,644$) of the addresses only focus on mint and burn transactions. We also find that 80\% of the investors have less than 15 Uniswap transactions, suggesting most of them are inexperienced on Uniswap. 
Nevertheless, we observe that many investors have thousands of transactions. For example, the address \texttt{0x80c5e69083}\footnote{0x80c5e6908368cb9db503ba968d7ec5a565bfb389} has the most mint transactions, and it has added liquidity $18,276$ times on 209 pairs. 
We further analyze the interacted liquidity pools of these participants. 
Roughly 45\% ($245,417$) addresses have transactions with only one pool, and over 90\% ($494,755$) of the addresses have transactions with less than 15 pools. However, 27 addresses have interacted with over 1K pools.
We manually inspect them and find that they are likely to be trading bot contracts engaged in arbitrage activities due to their repeated trading behaviors.

\subsection{Summary}
Uniswap has attracted a large number of tokens and created a prosperous trading environment. Nevertheless, there are many liquidity pools that were not created for long-term uses, since they have a low level of liquidity and only a few users joined in the trading activities of these pools. 
Most of the participants are new to Uniswap. The explosion of DeFi projects may attract these inexperienced investors, which can also be exploited by attackers. 
\begin{figure}[ht]
\centering
\includegraphics[width=0.99\linewidth]{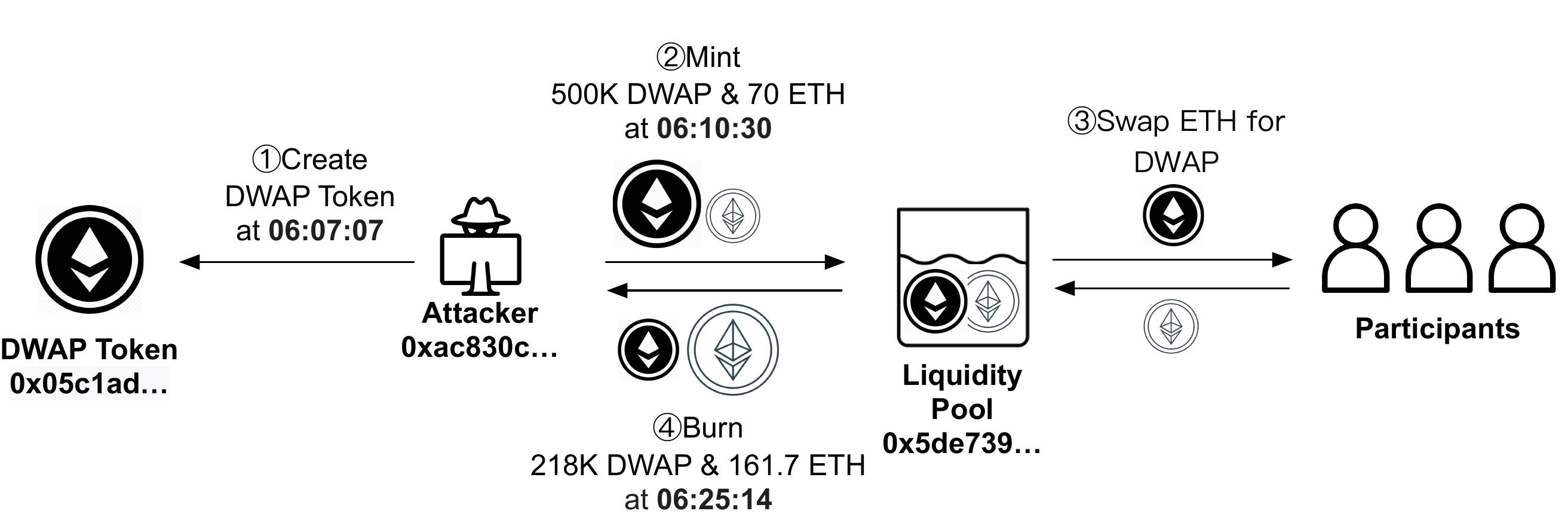}
\caption{A motivating example of a scam token \texttt{Deriswap (DWAP)}.}
\label{fig:example}
\end{figure}

\section{Identifying Scam Tokens on Uniswap}
\label{sec:detection}

\subsection{A Motivating Example}
As a decentralized exchange, Uniswap does not enforce any rules for token listing, i.e., anyone can list a token and create a liquidity pool freely. Thus, scammers can take the opportunity to list scam tokens to cheat unsuspecting users.
In this study, we find that there are many tokens with same/similar names, which are highly suspicious to be scams. More and more evidence shows that scam tokens have appeared on Uniswap.
For example, on Nov 23th 2020, shortly after Andre Cronje, the famous Yearn Finance (YFI) creator, announced his new DeFi project Deriswap, attackers created a fake token \texttt{Deriswap (DWAP)}\footnote{Token address:0x05c1ad0323b3f7f25cff48067fa60fa75dc7ba4f} and a liquidity pool had been created on Uniswap~\cite{news30min}. The whole process of this scam is shown in Figure~\ref{fig:example}. The scammer\footnote{Address:0xac830c76fc37ef3dd4c28c9b7ee548d1a46112eb} adds liquidity with 70 ETH and 500K DWAP tokens initially, and later removes liquidity of 217K scam tokens and 161.7 ETH. The attacker profits roughly 90 ETH (roughly \$54K)\footnote{Since the prices of tokens fluctuate every day, we calculate the profit of these scam tokens according to the price of top 50 tokens on December 6th 2020 on Uniswap. It applies to all remaining content.} considering the fee and swap cost. It is very surprising to see that, the whole process from the token creation to the withdrawal only took under 20 minutes. It is a common type of scam called ``Rug Pull'' in Uniswap, which will be detailed in Section~\ref{sec:analysis}.
This makes us wonder how many scam tokens/pools are listed and to what extent they have an impact on the investors. Thus, in the following, we develop a reliable approach to identify scam tokens/pools on Uniswap, and further characterize them.

\begin{figure}[ht]
\centering
\includegraphics[width=0.99\textwidth]{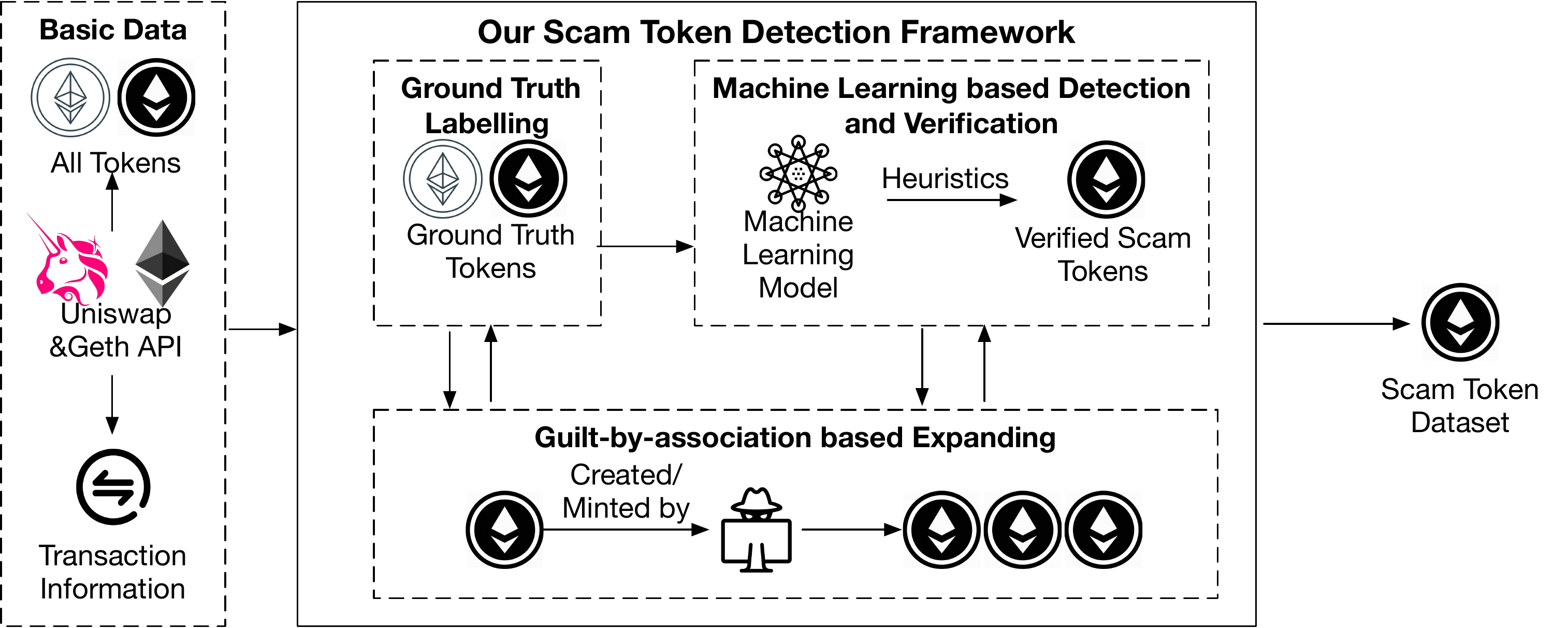}
\caption{The workflow of detecting scam tokens on Uniswap.}
\label{fig:structure}
\end{figure}

\subsection{Approach Overview}

\subsubsection{Key Idea}
Our preliminary exploration suggests that scammers usually list tokens and pools that look very similar to the existing cryptocurrency projects, due to the less regulation of both Uniswap and Ethereum.
The targeted projects are usually official tokens (e.g., USDT) that have been already released on Uniswap, or famous DeFi projects that are looking to conduct a token sale. 
Further, the scam pools are usually \textit{short-lived}, as the scammers would remove liquidity soon when there are victims falling into the traps. It suggests that the scam tokens and liquidity pools have quite unique features when compared with other normal tokens/pools, and these features can be used to distinguish scam tokens from the normal ones. 

\subsubsection{Overview of our approach.}
The overall workflow of our scam detection framework is shown in Figure~\ref{fig:structure}, which is made up of three major components.
(1) \textit{Ground truth labelling component} is used to collect official (normal) tokens and the most reliable scam tokens (i.e., the fake tokens whose names or symbols are identical with the official ones). The labelled ground truth dataset is used as the seeds for further expansion.
(2) \textit{Guilt-by-association based expansion component} is used to enlarge our labelled scam token dataset based on two reliable heuristics. 
(3) \textit{Machine learning based scam detection and verification component} is used to identify more scam tokens based on the features learnt from our labelled dataset. To eliminate potential false positives, we use a strict verification strategy to only label the most reliable scam tokens. 
Note that, guilt-by-association based expansion would be further applied to the identified new scam tokens, and produce our final results. 
Our approach considers both the naming characteristics and the transaction behaviors of scam tokens. Further, our approach is powered by the strict verification strategy, and thus it can produce the most reliable results.

\subsection{Ground truth labelling}
\label{sec:tokencollection}

\subsubsection{Official tokens.} 
We first collect a list of popular tokens from the CoinMarketCap~\cite{coinmarketcap} ranking list, and the Etherscan~\cite{etherscan} ranking list, and then use the following method to manually verify them. The popular official tokens usually have been listed on large CEXs (i.e., they have been verified by the operators of CEXs), with the exchange rates for US dollars.
Since some official tokens may migrate from old addresses to new addresses (e.g., due to security issues), we further flag those old token addresses as official ones too.
Through this way, we collect $2,397$ official tokens in total.

\subsubsection{Scam tokens.} 
We label the scam token seeds in two ways. 
First, as Ethereum does not enforce any restrictions on the names and the symbols of the newly created tokens, some fake tokens use \textit{identical} identifier names to imitate the official tokens to trick victims by means of airdrop scam and arbitrage scam~\cite{gao2020tracking}.
By comparing the token names and the symbols of all the ERC-20 tokens in Uniswap with the labelled official tokens, we have flagged $4,017$ fake tokens. 
Further, as Etherscan usually marks phishing or scam tokens, we implement a crawler to collect the tags of tokens, which collects $31$ more scam tokens.

\textit{In total, our ground truth labelling phase has collected $6,445$ tokens in total, with $2,397$ official tokens and $4,048$ (=$4,017+31$) scam tokens.}

\subsection{Guilt-by-association based expansion}
\label{sec:expand}

Empirically, scammers usually create more than one scam tokens to expand the scale of their scam campaigns. Therefore, we mark all Ethereum accounts that have created a scam token or a liquidity pool (i.e., the pool that trades scam tokens) in our labelled dataset as \textit{scam creators}.
For other tokens/pools created by these scam creators, they are highly suspicious to be scam tokens as well.
We call this strategy ``Guilt-by-association'', which has been used in previous work to identify malicious domains and malware~\cite{sebastian2020towards,khalil2018domain}. In particular, this strategy could also help us find new emerging scam tokens released by the same scam attack campaigns, even with no users fallen into the scams yet (i.e., no or few transactions). 

\subsubsection{Expansion based on scam token creators.} 
After excluding 7 addresses that are tagged by Etherscan as \textit{Contract Deployer}\footnote{The Contract Deployer can be used by different users to create contracts with similar functionality or structure, thus these addresses should be excluded when expansion}, $2,972$ \textit{scam creators} are marked. 
We then identify all the tokens created by these scam creators and obtain $2,424$ new candidate scam tokens.
We then manually verify a portion of popular tokens (i.e., have more than $1,000$ transactions) and less popular tokens (i.e., have less than $1,00$ transactions) 
to verify the reliability.
Specifically, we search the tokens on Google to check whether the official websites exist, and whether there are scam accusations on BitcoinTalk and other forums, etc.
Among $2,424$ candidates, 28 of them have more than $1,000$ transactions. We have manually verified these 28 popular tokens alongside 50 tokens with few transactions (10 to 100) and find no false positives, which suggests the reliability of our heuristics. 

\subsubsection{Expansion based on scam pool's creator and first mintor} 
The Ethereum account which creates the scam liquidity pool (i.e., the liquidity pool reserves a pair consisting of the scam token and other token) 
and \textit{firstly} adds liquidity to the scam pool is marked as a scam creator as well. 
In this way, we flag $2,985$ scam creators.
And $2,609$ (87.4\%) of them are overlapped with the \textit{scam token creators} labeled in the previous step.
We further find the tokens created by these scam addresses and expand $24$ new scam tokens through this method. Similarly, we have manually verified these 24 tokens and find no false positives.

\textit{In total, based on the labelled scam token seeds, we further expand $2,448$ ($=2,424+24$) scam tokens. These tokens, along with the labelled tokens in Section~\ref{sec:tokencollection}, will be used to train a machine learning classifier for identifying more scam tokens. Note that, we will further use reliable heuristics to verify the scam tokens flagged by the machine learning classifier, and the expansion method will be further adopted to new confirmed scam tokens to enlarge our dataset (see Section~\ref{sec:tokendetection}).}

\subsection{Machine learning based detection and verification} 
\label{sec:tokendetection} 

The aforementioned two phases can only flag the most obvious scam tokens. However, there are many other scam tokens that impersonate token sales for popular DeFi projects (e.g., Teller Finance) and famous brand names (e.g., Facebook). It is non-trivial for us to get a list of targeted tokens/DeFi/brands for comparison. Thus, in the following, we seek to identify scam tokens based on their transaction behaviors on Uniswap.
We first train a machine learning classifier, and apply it to all the unlabelled tokens (see Section~\ref{sub:classifier}). For the flagged suspicious tokens, we examine them and summarize several highly reliable heuristics for verification (see Section~\ref{sub:verification}). Finally, for the newly verified scam tokens, we further adopt the expansion technique (see Section~\ref{sub:expansion}). 

\subsubsection{Machine learning classifier.}
\label{sub:classifier}

Based on our preliminary observations, we use a comprehensive set of features to train a scam token classifier (see Table~\ref{tab:feature} in Appendix).

1) \textit{Time-series features}. 
Scam tokens and pools are usually short-lived. Once the scam has attracted some victims, scammers tend to remove all liquidity of the pool to get all the reserved tokens in the pool and gain a profit. 
The scam token and the corresponding liquidity pools would be discarded quickly when attackers succeed in scamming money, as it is easy for the scammer to launch a new scam token. Thus, for each token, we analyze its \textit{active period} (i.e., from its first transaction to the latest one) and use it as one feature. Note that, to eliminate the potential bias introduced by our dataset collection process (e.g., a new normal token listed on Uniswap would lead to a short \textit{active period} in our dataset), we further consider the \textit{active interval} between the last transaction of the token and the time of our dataset collection, as a feature.

Further, we observe that the distribution of transaction events (i.e., mint, swap, burn) in scam tokens is quite different from the normal ones. For example, for a scam token, the mint events are more concentrated at the beginning of its life-cycle (i.e., scammers provide the liquidity to attract victims), while the burn events are more concentrated at the end of its life-cycle (i.e., scammers remove the liquidity to gain a profit). As a contrast, for official tokens, mint and burn events are distributed across the life-cycle.
Thus, for a given token, we propose to analyze the relative position of different types of  events (i.e., mint, swap, burn) in terms of occurrence time, which can reflect the activity of a token to some extent,
and define the relative time position of each type of event as: 
\begin{equation}
\label{eqn:timeposition}    
P_{event}= \frac{\frac{1}{n}\sum_{i=1}^n (T_{i}-T_{start})}{T_{end} - T_{start}},
\end{equation}
where $n$ is the number of events of the specific type (i.e., mint, swap or burn), $T_{i}$ denotes the timestamp of the $i$-th event, $T_{start}$ and $T_{end}$ represent the timestamps of this token's first and last transactions, respectively. 
If a token's mint events are concentrated at the initial stage, then $P_{mint}$ will be close to 0. For a scam token, the $T_{end}$ is highly likely to be the time when its only burn event initiated by the attacker, thus the $P_{burn}$ of this token will approach 1. 
Thus, we further extract five features related to the time position of each type of event, including mint events, swap events, swap-from events (i.e., swap target token for the other token), swap-to events (i.e., swap other token for target token) and burn events.
In total, we extract seven time-series features, which are shown in Table~\ref{tab:feature} of the Appendix.

2) \textit{Transaction features.}
The number of transactions can reflect the popularity and the volume of a token. 
Here, for a given token, we consider its transactions on both Uniswap and the overall Ethereum network. A trustworthy official token will have transactions beyond Uniswap. 
The extracted transaction features include the total number of transactions on Uniswap and Ethereum respectively, the number and the proportion of the transaction events (i.e., mint, burn and swap) and the number of involved addresses, with 24 kinds of features in total. 

3) \textit{Investor features.}
We observe that a large portion of the investors of scam tokens are inexperienced, i.e., associated with few transactions. Thus, we speculate that the greedy newcomers are the major targets of the scammers. Thus we extract 4 kinds of investor features for each token, including the average number of trading pools they interacted with (mint/burn or swap), and the average number of (mint/burn or swap) transactions.

4) \textit{Uniswap specific features.} 
For each token, we further extract features from its state on Uniswap, including the number of liquidity pools it involved, its trade volume, the total liquidity of the token, etc. 
The details of these features are shown in Table~\ref{tab:feature} of the Appendix. 

Based on the extracted features, we next train a machine learning classifier. We have tried different kinds of models to train the classifier, including Logistic Regression~\cite{dreiseitl2002logistic}, SVM~\cite{chang2011libsvm}, Random Forest~\cite{breiman2001random}, and XGBoost~\cite{chen2016xgboost}. 
We use the 10-fold cross validation to evaluate these models. We shuffle our dataset randomly and split the dataset into 10 groups. For each group, we take this group as a training set and take the remaining groups as the test set to evaluate the model. Our experiment results suggest that the random forest model achieves the best result. Using the random forest model, the Precision, Recall, and F1 score of our classifier are 96.45\%, 96.79\% ,and 96.62\%, respectively. It suggests the high accuracy of our approach. Thus, we further apply the trained classifier to the unlabelled tokens on Uniswap. Our classifier flags $11,182$ tokens as potential scam tokens based on their transaction behaviors. 

\subsubsection{Verification.}
\label{sub:verification}

Our machine learning classifier flags the suspicious tokens with potential scam behaviors. Although our classifier can achieve excellent results, it cannot achieve 100\% accuracy. As our goal is to characterize and measure the landscape of scam tokens on Uniswap, a dataset with high accuracy is a must. Thus, we next use a strict verification strategy to label the most reliable scam tokens, and perform the characterization study based on them.
We randomly select 200 flagged suspicious tokens, seeking to find the clues that can be used to confirm they are scams. By analyzing their token names/symbols, we devise two highly reliable rules. 

First, we find that many of the flagged tokens share the identical token names with each other. Although they did not counterfeit the popular official tokens we labelled in Section~\ref{sec:tokencollection}, we observe that they seek to promote the scams by exploiting the eye-catching DeFi projects and related hot topics. For example, we find that there are $429$ scam tokens that share the identical name and they pretend to be the famous DeFi project {\tt yearn.finance}~\cite{yearn} (YFI). As another example, there are 12 tokens named \texttt{bore.finance} and none of them are the official tokens since the real one~\cite{boretoken} is a \texttt{BSC} (Binance Smart Chain)~\cite{binancesc} token, rather than an ERC-20 token on Ethereum. Thus, we group the suspicious tokens based on their token names and symbols. For the suspicious tokens with identical names, we further search these names to verify whether they are counterfeit ones. In this way, $3,434$ of the suspicious tokens are flagged as scam tokens by us with high confidence.

Second, we observe that many scam tokens impersonate to be the tokens released by some popular companies, authorities, organizations or celebrities by using similar names of them, while actually there are no official tokens released by these entities. For example, we find a number of scam tokens with names related to Google, Amazon, TikTok, Trump, Elon Musk, etc. Thus, following the general way of manual labelling, two authors go over the token names and cross checked the results to eliminate the bias. We consider a token to be a scam token if both of the two authors label it as a scam. In this way, we identify $316$ such cases.

In total, by applying the heuristics to the suspicious tokens flagged by the machine learning classifier, we identify $3,750$ ($=3,434+316$) scam tokens. 
For the remaining $7,432$ (=$11,182-3,750$) suspicious tokens flagged by our classifier, we sample some tokens to manually inspect them and find that they are likely to be scams too, although we lack strong evidence. Some of these tokens have the name patterns of DeFi projects (e.g., ``finance'' or ``network'' keyword in their names) while we cannot find any official sites of them but some scam accusations in search engines or forums. Furthermore, some of these tokens share a similar trading behavior with the scam tokens we verified. For example, the {\tt Phoenix.Finance} (PF) token\footnote{Token address:0x03c2f1b1ba5c5a6cf6d0af816a721a5827171704} is suspicious since it only has 12 token transfer transactions and the creator of the token adds and removes liquidity within one day. The \texttt{Axonic Network} (AXO) token\footnote{Token address:0x0e123e061cf206dd2bb2c6e040e5932a7a6111e9} and \texttt{Shrink Finance} (SRK)\footnote{Token address:0x099c4772ac7c866c48b5d870f2c702193cd3f27d} also have the similar behavior and their corresponding addresses (i.e., token address, creator address, etc.) are accused of scam by users. However, they did not fall into the aforementioned two strict heuristics we proposed, and we did not incorporate them into our scam dataset, due to the conservative consideration.

\subsubsection{Expansion.}
\label{sub:expansion}

Following the ``guilt-by-association'' expansion, we expand our scam token list by analyzing the creators of newly identified $3,750$ scam tokens and liquidity pools, and obtain $674$ more scam tokens. Thus, in the machine learning phase, we identify $4,424$ scam tokens in total.

\subsection{Summary}
Based on the extensive analysis, we flag $10,920$ scam tokens with very high confidence, i.e., 4,048 scams flagged in the ground truth labelling phase, 3,122 ($=2,448 + 674$) ones are expanded based on guilt-by-association, and 3,750 tokens are expanded by using our machine-learning based detection and verification technique. These scam tokens are associated with $11,215$ liquidity pools.
We want to reemphasize that, \textit{the number of identified scam tokens is indeed a lower-bound, as we enforce a strict verification method to get the most reliable results}.

\section{Characterizing the Scams}
\label{sec:analysis}

We next characterize the flagged scam tokens and liquidity pools by investigating their scam behaviors, the scammers, and the financial impact.

\subsection{General Overview of Scam Tokens and Scam Liquidity Pools}

\subsubsection{Scale} 
We have identified a total of $10,920$ scam tokens with $11,215$ liquidity pools, accounting for 50.14\% of all the tokens (44.63\% of all pools) on Uniswap. Their total trade volume reaches over \$365 million. $88,567$ Ethereum addresses have interacted with these scam tokens with $659,087$ transaction events in total. \textit{It suggests that Uniswap is flooded with scam tokens}. 

\subsubsection{The trend of scam tokens}
\begin{figure*}[t]\centering 
\subfigure[Token Listing and Pool Creation.]{ 
\begin{minipage}{0.3\textwidth}\centering
\includegraphics[width = 0.9\textwidth]{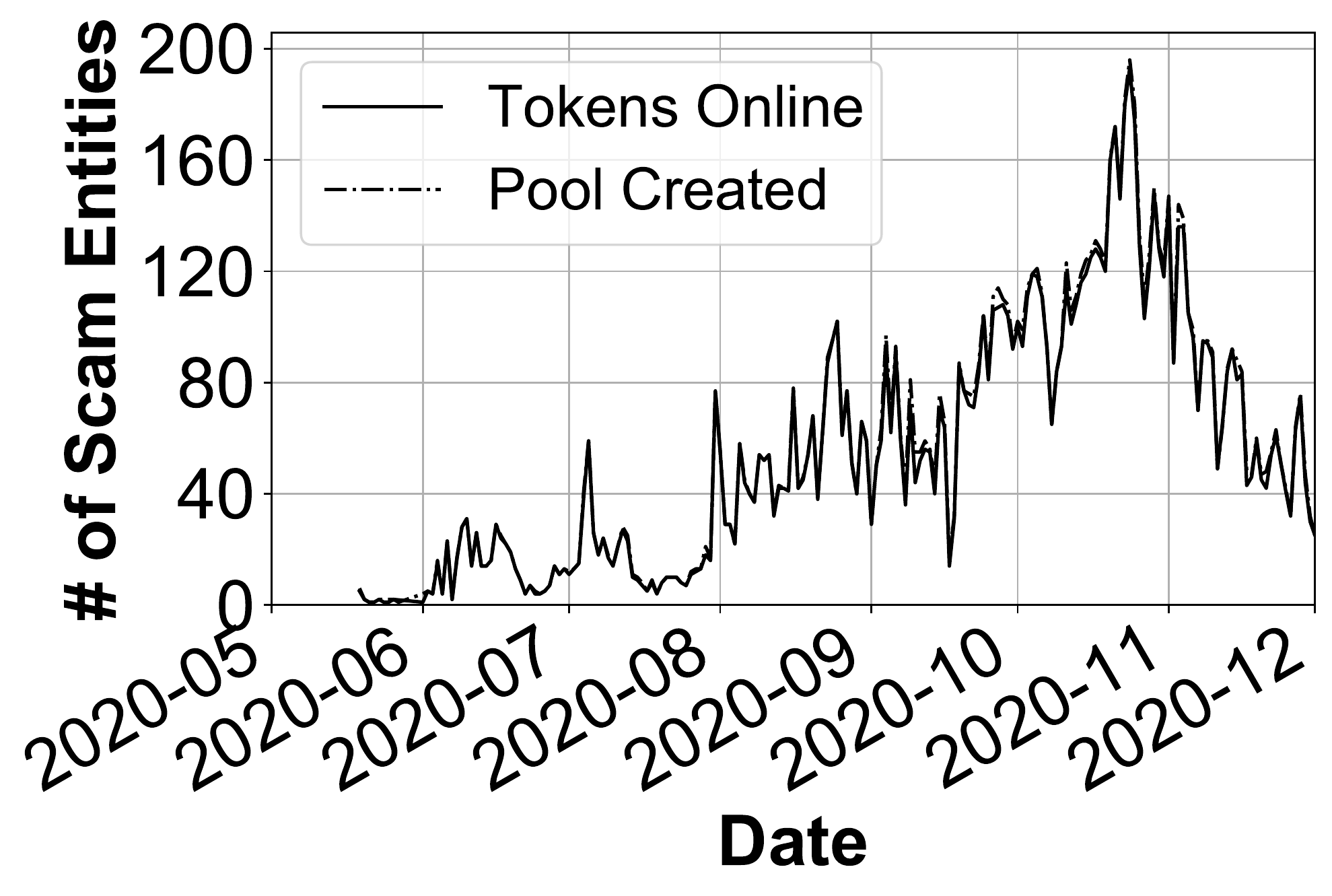} 
\end{minipage}}
\subfigure[Transaction Events]{ 
\begin{minipage}{0.3\textwidth}
\centering 
\includegraphics[width = 0.9\textwidth]{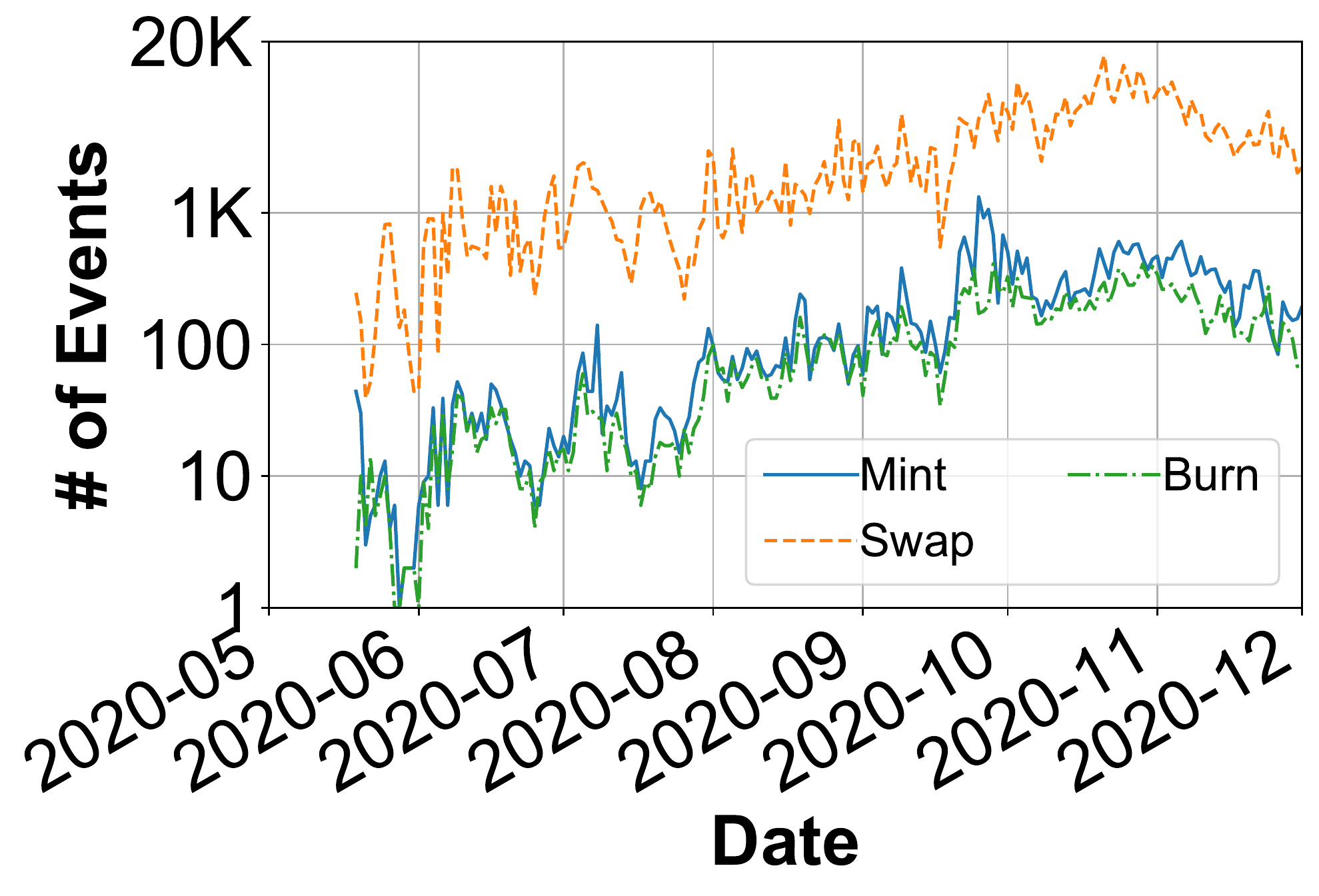} 
\end{minipage}
}
\subfigure[Trade Volume and Liquidity]{ 
\begin{minipage}{0.3\textwidth}
\centering 
\includegraphics[width = 0.9\textwidth]{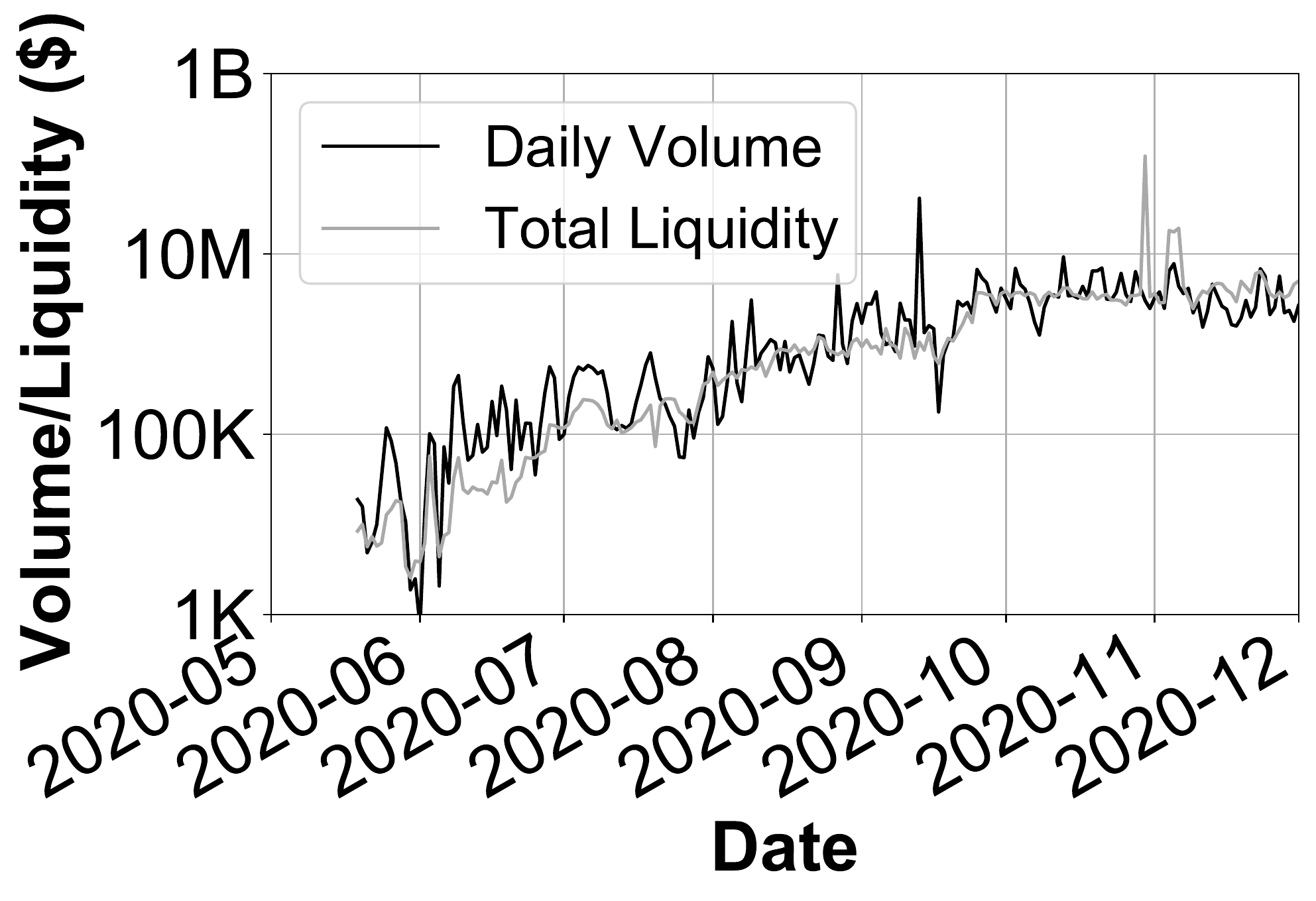} 
\end{minipage}
}
\caption{The trend of scam tokens and liquidity pools on Uniswap.}
\label{fig:scam_evo}
\end{figure*}

As shown in Figure~\ref{fig:scam_evo} (a), the creation of scam tokens and scam liquidity pools roughly follow the overall trend of Uniswap (see Section~\ref{sec:uniswap_evo}) and the peak appeared in October 2020, where over 3.8K scam liquidity pools were created on Uniswap. 
In our dataset, the first scam token \texttt{Bizcoin} (BIZ) (which is flagged by heuristics in Section~\ref{sub:verification} and there are 2 tokens with name ``Bizcoin'' and 3 tokens with the symbol ``BIZ'')\footnote{LP token address:0xde65eed30da8107ce49e8f1952391e16756c2998} appeared on May 19th 2020 and there were four other scam tokens listed on Uniswap this day.
According to the post on 4chan~\cite{biz_4chan}, it was promoted as a community token and appeared to be profitable. The token received over 1K transactions and earned the creator about \$2200. 
Figure~\ref{fig:scam_evo} (b) shows the trend of scam tokens' transaction events on Uniswap. 
On average, there are over $3,200$ transaction events related to scam tokens daily, and the peak reaches almost 20K transaction events. 
The volume and the liquidity of scam tokens are shown in Figure~\ref{fig:scam_evo} (c)\footnote{The daily volume and liquidity data of scam tokens come from Uniswap API}. 
On average, scam tokens have a daily trade volume of \$1.8 million. Different from the general trend, the trade volume of scam tokens often exceeds their liquidity, indicating the skyrocketing in price within a short time. \textit{The scammers often use this trick to attract more people to invest in the scam tokens} (see Section~\ref{sec:behave}). 

Besides, we also compare the creation times of scam tokens on Ethereum and their corresponding pools on Uniswap, to investigate if these tokens are organized in scam campaigns. 
In our dataset, $10,770$ (98.6\%) tokens were created after the Uniswap V2 launch and $5,835$ (53\%) scam tokens were created in September and October, which was the most active period of Uniswap. Besides, for over 92\% of the scam tokens, their creation on Uniswap and their related liquidity pool's creation on Uniswap were done within one day. This suggests that \textit{most of the scam tokens were created specialized for carrying out scam campaigns on Uniswap.}

\subsubsection{The pools of scam tokens.}
\label{subsec:scampool}
Roughly 98\% ($10,686$) of the scam tokens have only 1 scam liquidity pool. 
Nevertheless, some attackers try to create a number of pools to reach as many victims as possible. For example, the token \texttt{LEV}\footnote{Address:0x3868bd6e8b392eb8dbc8cdcd0c538dc66529adbe} has been paired with 11 kinds of tokens, among which 10 tokens are the leading official tokens and the other 1 scam token was minted by the same liquidity provider of \texttt{LEV}. 
Besides WETH, other official tokens like USDT, USDC and DAI are also favoured by attackers. 
We further investigate how long it takes for the scammers to remove the liquidity they inserted (i.e., usually after victims rushed to the scam pools)
by calculating the interval between the scammers' first mint and burn events. 
Surprisingly, over 86\% of the scam liquidity pools have an interval within 1 day, and 37\% of the pools' liquidity were removed within 1 hour.
\textit{This suggests that attackers prefer to act quickly to secure their scammed money before the victims took actions.}

\subsection{Understanding the Scam Behaviors}
\label{sec:behave}
We next investigate the behaviors of these scam tokens and liquidity pools, i.e., how they cheat unsuspecting users and get a profit. We first randomly select 100 scam liquidity pools, and manually examine their transactions on Uniswap and their corresponding token smart contracts to investigate their scam behaviors. Then we design methods to check all the liquidity pools and scam tokens in our dataset.
In general, all the scam liquidity pools are created for the ``rug pull'' scams, while some of the scam tokens use many tricks to secure or enlarge the scam tokens' profits. 

\subsubsection{The ``Rug Pull'' Scams.} 
A rug pull is a common kind of scam where developers abandon a project and take their investors' money~\cite{rugpull}. 
From the liquidity pools' perspective on Uniswap, the ultimate purpose for performing rug pull scams is to fool the victims to invest in the scam tokens they created and then drain the money of the pools. 
The motivating example in Figure~\ref{fig:example} shows a case of the rug pull scam.
The scammer usually creates a scam token and then provides liquidity of the token by pairing it with a leading cryptocurrency on Uniswap. They will promote the scam token through social networks with attractive advertisements, usually through Telegram. 
When enough victims rush into the liquidity pool and exchange for the worthless tokens with valuable WETHs or other stable coins, the scammer will withdraw everything from the liquidity pool, and the victims will get nothing but the worthless scam tokens instead. \textit{It can explain why many liquidity pools on Uniswap have low levels of liquidity, as we observed in Section~\ref{sec:poppair}.}
By analyzing the mint and the burn transaction events, all the scam liquidity pools are carrying out the ``Rug Pull'' scams.

Further, the ``Rug Pull'' scams are often combined with other tricks. 
Most of the attackers were found to swap scam tokens using tricks like \textit{pump-and-dump scams}~\cite{pumpdump}, making a scam coin skyrocketing in price within hours. 
Due to the mechanism of Uniswap, purchasing tokens will raise the price and the volume of these tokens. \textit{This trick will create the illusion that the scam token is popular and profitable, which can attract inexperienced investors.} Since attackers can sell the scam tokens they have or remove the liquidity of the scam token pools after the rise of token prices, most of the money they invested in the scam pool will eventually go back to the attackers and many attackers are willing to perform this trick. In our dataset, 93\% of the liquidity pools have ever had swap transactions initiated by token/pool creators or their collusion addresses, and some even swapped for a large amount of money. For example, the token creator 0x2faea647a49a43187ff19cdd5698489ea9a6acb1 swapped 150 WETHs for the token \texttt{RadixDLT.com (RADIX)}\footnote{Token address:0x4b7266fa8ffda838c64c4b93a7092afe4bd68ed4} , which leads to the increase of the price by about 80\%. 
Besides using the creators' addresses to add liquidity and swap for tokens, many scam campaigns are using multiple collusion addresses to add/remove liquidity or swap tokens. These addresses could be operated like normal investors, which makes them hard to be detected.

\textbf{Second-round Scams.}
Besides, 433 scammer addresses are found to perform second-round scams, i.e., after they make a profit by removing liquidity from scam token pools, scammers add liquidity again to the same pools and start a new round of scams. 
For example, the pool creator \texttt{0x1bf3bd8e8afe80d786caa69f98385e0aa7e312ff} created liquidity pool for Xfinances (XFIS) token\footnote{Token address:0xac51e84ccf9ff013f54cc53bbed80250e558d1aa} and added liquidity to it on 2020 October 9th. He earned over 72 ETHs (roughly \$43K) through the rug pull scam in about 30 minutes. And only after 4 minutes, he added liquidity again. He then removed liquidity again after 6 hours, and made another profit of 26 ETHs (roughly \$16K). \textit{It suggests that many victims never check the transaction history of the pool before they rush into it.}

\begin{figure}[t]
    \centering
    \includegraphics[width=0.99\linewidth]{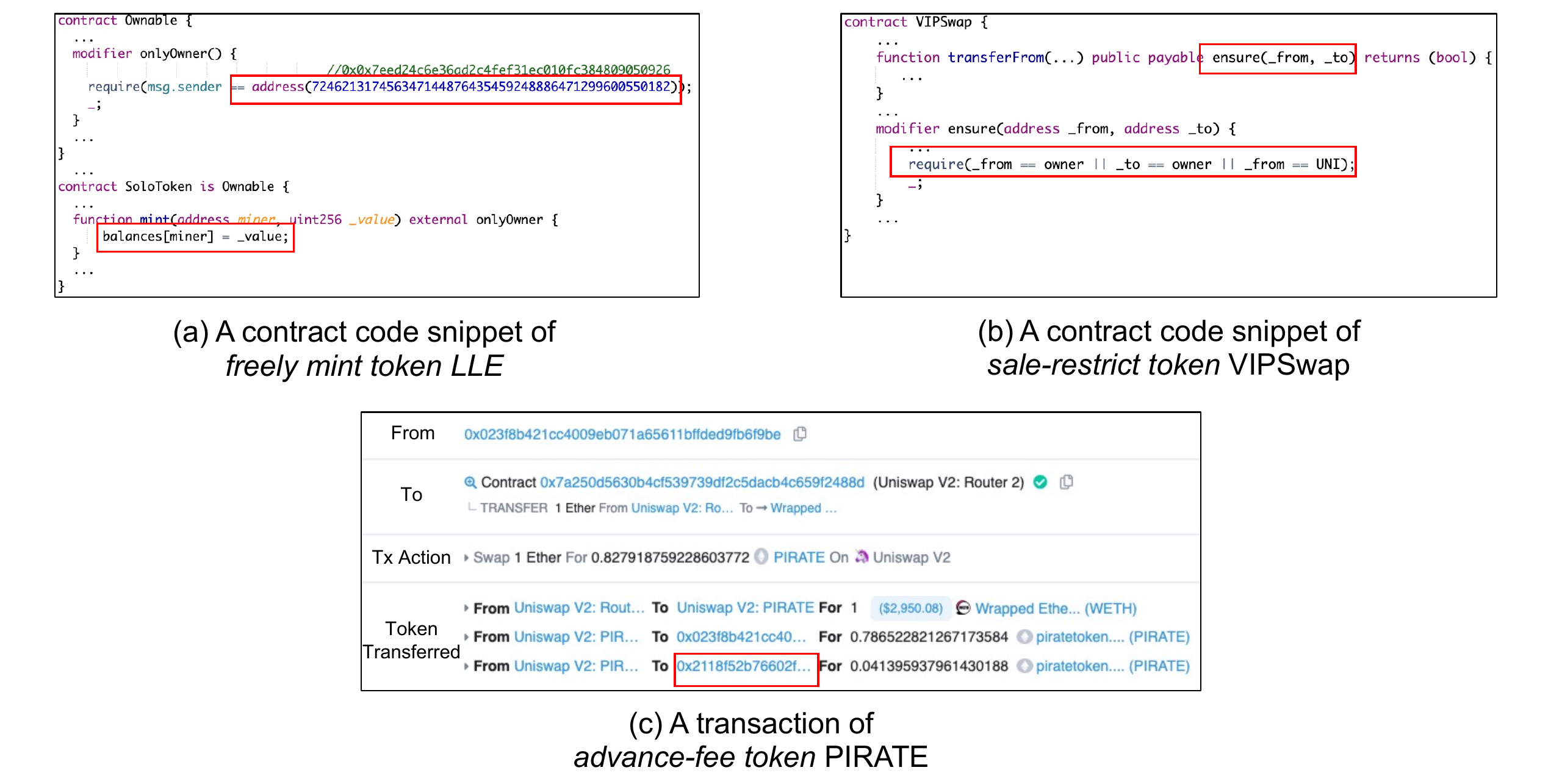}
    \caption{Examples of token contracts with backdoors and advance-fee tokens.}
    \label{fig:trap}
\end{figure}

\subsubsection{Scam Token Smart Contracts with Traps.} 
From the perspective of scam tokens, some of them have inserted well-designed traps in the code to further cheat the investors. 
Through exploring abnormal transactions involving multiple recipients and investigating open-source scam contracts, we have observed two types of such tricks and designed a method to identify them.

\textbf{Backdoors in scam token contracts.}
In general, a scam token campaign would meet two issues in successfully carrying out the scams. 
First, some experienced investors may not believe in the token/pool creators who hold most of the tokens they issued. 
Second, once victims found they were cheated after the trading in Uniswap, they usually seek to mitigate a loss, i.e., by swapping back their valuable tokens as soon as possible before the liquidity has been removed by the scammers.
\textit{Thus, we observe that some advanced actions could be performed by scammers to either prevent the victims from getting back their money or gain the trust of investors.}

For the first issue, token creators could design malicious \textit{freely mint token contracts} to enable specific scam addresses to add their token balance deliberately while claiming they do not hold many scam tokens to reduce the victims' suspicions. By doing so, these specific addresses could swap the tokens in the liquidity pool when the token price is at a high point (i.e., some victims rushed to the pool). These specific scam addresses could perform this kind of operation stealthily since this kind of operation may even not emit events for users to track based on the design. Figure~\ref{fig:trap} (a) shows an example of this kind of scam token contract, where the \texttt{Leopard lending ecology} (LLE) token \footnote{Token address:0x48fa649638318aa0e85dc0fec425c015304d175a, its contract name is called ``SoloToken''} left a back door for the scam address \texttt{0x7eed24c6e36ad2c4fef31ec010fc384809050926}. The address can call the ``mint'' function to add any amount of LLE token for it deliberately. The function was not implemented to emit an event and thus users and Dapps rely on events to track the transfers of this token cannot find the mint activity easily. In fact, the address added $1.8\times10^{12}$ tokens to it (which are even more than the token's total supply) and swapped these $1.8\times10^{12}$ tokens to get roughly $474$ Ethers (roughly \$28K) from the LLE liquidity pool.

To address the second issue, some token creators come up with a trick to prevent victims from selling tokens except for token creators themselves by designing \textit{sale-restrict token contracts}. Figure~\ref{fig:trap} (b) shows such an example. 
As written in the contract modifier, the \texttt{VIPswap} (VIP) token~\footnote{Token address:0x3b0407c648dd2f3eaa23fc69f952d98b2f24257e} allows all the users to buy VIP tokens while restrict all the users except the contract owner to sell them. 
The token creator \texttt{0x7af0f3e99a30b682d61c07be19c5874fb80e3832} created 58 such tokens and these tokens gained a profit of over $77$ Ethers (roughly \$46K).

To identify the backdoors, we first analyze the token contracts and identify the contracts that used \textit{Solidity modifiers} in their key functions (like selling tokens or minting tokens), as the Solidity modifier is mainly used for automatically checking a condition prior to executing a function.
Then, we manually check the code to see if these modifiers are used to restrict the functions to be executed only by attackers. 
By this, we find 297 \textit{freely mint token contracts} with 131 token creators and 373 \textit{sale-restrict token contracts} token contracts with 109 token creators.

\textbf{Advance-fee tokens} 
Besides gaining a profit from ``Rug Pulls'' scams, some attackers have designed tokens that will charge a fee when users perform mint, swap, or burn operations. 
For an \textit{advance-fee token}, the fee rules are often written in its codes to transfer part of the tokens to a specific address every time when a swap operation happens. This specific address is claimed to be a bonus or reward pool to reward the users involved in the transactions of this token.

To identify them, we track all the transactions related to scam tokens, and identify abnormal transactions where scammers received tokens or ETHs even though they are not the direct participants of these transactions. 
We have identified 63 advance-fee tokens.
A typical example is shown in Figure~\ref{fig:trap} (c)\footnote{Transaction Hash:0xb63891bebda1d330e06603d680accfb1c5ace82c1e473e5cfa620b511d544ae9}. The \texttt{piratetoken.finance} (PIRATE) token\footnote{Token address:0x94152edd72eab86c016c3c5fb40376a88f10de5b} will charge a 5\% fee from investors to the address \texttt{0x2118f52b76602fde203f4ad1ea48690223af7568}, which is declared by the scammer as a daily bonus to a random user address. However, the address will never send out the bonus and this is just an excuse to stimulate token transfers and intend to raise the token's price in disguise.

\subsection{Understanding the Scammers}
\label{sec:scammers}

As aforementioned, different kinds of scam addresses controlled by the scammer would collude to carry out a scam. In general, the following five kinds of scam Ethereum addresses are involved:
1) \textit{scam token addresses}, the token used to carry out scams; 
2) \textit{scam liquidity pool addresses}, the liquidity pools consist of pairs of scam tokens and other tokens;
3) \textit{the creators of scam tokens}, addresses that create the scam tokens on Ethereum; 
4) \textit{scam pool creators (first mintors)}, addresses that create the scam liquidity pools on Uniswap; 
and 
5) \textit{collusion addresses}, addresses which cooperate with scam token/pool creators to carry out scam campaigns. 
Since scam tokens and scam pools have already been investigated in previous sections, we will analyze in detail the scam token creator addresses, scam pool creator addresses and collusion addresses.

\subsubsection{The creators of scam tokens.} 
The $10,920$ scam tokens are created by $6,288$ scammers, and 89\% of them were first minted by their creators. 
Roughly 76\% of the scam token creators only created one scam token while 1\% of the scam creators have created more than 10 scam tokens. The scam creator that released most number of tokens is \texttt{0x3bcfa9357ab84baec04313650d0eebb3fd51070d}, with 87 scam tokens in total, including several counterfeit tokens like \texttt{Keep3r} (KPR)\footnote{Token Address:0x66f04254ca406cedf222687afe873a35da573f2c}, \texttt{YKeep3r.network} (YKP3R)\footnote{Token Address:0x7a1c0213c9e05ed1b20d691ceda7387a62725143} ,etc., targeting at \texttt{Keep3rV1} (KP3R)~\cite{keep3r}, a famous DeFi project.

\subsubsection{The creators of scam liquidity pools.} 
As to the liquidity pools, 11,215 scam pools are created by $6,465$ creators (first mintor).
\textit{When considering liquidity providing, pool creators usually initially provide a substantial amount of liquidity into their pool to cultivate investor confidence, as the liquidity they provided will eventually go back to them.} 
Over 60\% of the pool creators have ever provided liquidity with valuable tokens at a cost of more than \$10K. The most lavish pool creator has provided liquidity with $1,600$ ETH (roughly \$962K) in the {\tt Cybercore.Finance} (CYBER)\footnote{Token address:0x5f21f580261a773aab2bff6cbc9814f6e7a67d78}-WETH pool.

\subsubsection{Collusion Scam Addresses.}

To attract victims and avoid the scams being easily detected, collusion scam addresses are usually utilized to collaborate with the scam token/pool creators to carry out scams.
Some collusion addresses are participated in providing and removing liquidity of the scam pools while the other collusion addresses are used to swap tokens (e.g., like the pump and dump aforementioned). 
The collusion addresses could swap valuable tokens for scam tokens to raise the price of scam tokens, or in contrast, they could swap scam tokens for valuable tokens back to drain the pool and make a profit. 
In the case of \texttt{Super Core Reserve Token} (SCRT)\footnote{Token address:0x002ef27dee7a7d74ba59671385c51aa3d561d228}, the token creator and pool creator \texttt{0xc7f82560c727c2045e7c19f8bc29c5cb8d258f7c} first transferred 750 SCRT to its collusion address \texttt{0x39e407b5cf03311251c79c60dbc72b842007ba12}. Then this collusion address waited for the price rise, sold out the tokens it had and transferred the ETHs it got to the creator. \textit{The behaviors of collusion addresses may look similar to victims on Uniswap and it is hard to distinguish them based solely on Uniswap events, while we should not regard the collusion addresses as victims.} 
Thus, we further design a method to accurately detect collusion addresses.

\begin{figure}
    \centering
    \includegraphics[width=0.99\linewidth]{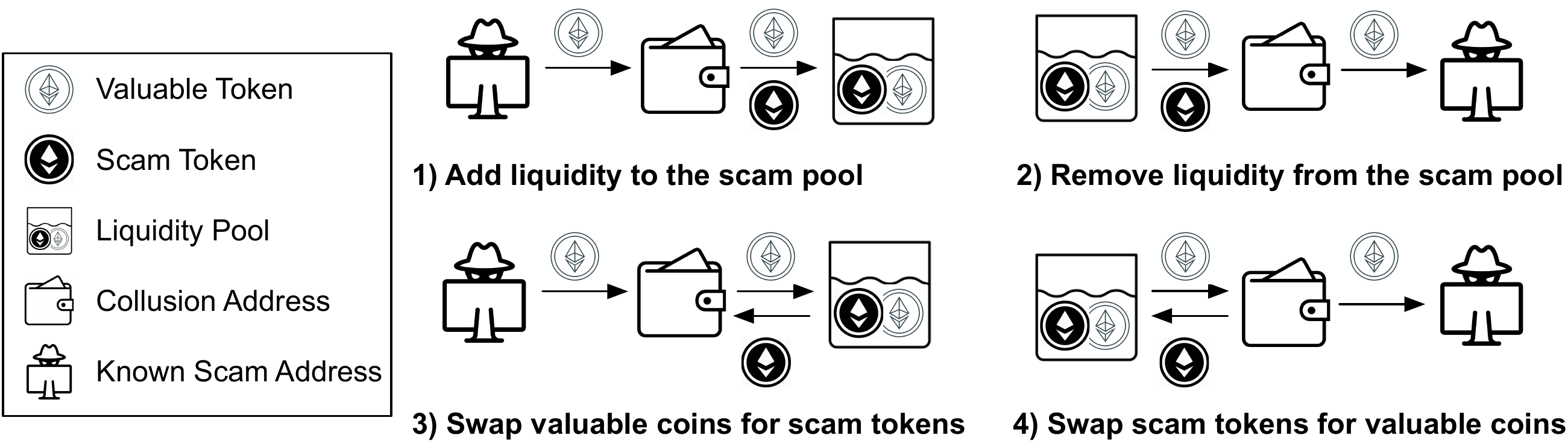}
    \caption{Four kinds of collusion addresses categorized based on their Uniswap transaction behaviors. The arrow denotes the transfer of tokens and the time order of transaction events is from left to right.}
    \label{fig:collusion}
\end{figure}

\textbf{Detecting collusion addresses.} 
One major characteristic of the collusion addresses is that \textit{they should have strong connections with other scam addresses operated by the same scammer}. They first need to operate on the same scam Uniswap pool with other scam addresses. Besides, there should be money flows between the collusion addresses and other known scam addresses, as the collusion addresses may either receive money from scammers to interact with the pools (i.e., add liquidity or swap scam tokens) or transferring money they earned (by removing liquidity or sell scam tokens) to the scammers for aggregation.
Thus, to effectively differentiate collusion addresses and victims, we have categorized the collusion addresses into the following four categories based on their transaction behaviors (i.e, mint, burn, and swap) on Uniswap and summarized their features, which is shown in Figure~\ref{fig:collusion}.

\begin{itemize}
    \item[1)] \textit{Add liquidity to the scam pool.} 
    For the addresses that inserted liquidity to the scam pool, if they have ever received Ether or stable coins (i.e., according to the corresponding token pairs of the pool) from the known scam addresses of the pool (e.g., scam token/pool creators) \textit{before} their adding liquidity transactions, we will flag them as collusion addresses.
    \item[2)] \textit{Remove liquidity from the scam pool.} For the addresses that removed liquidity from the scam pool, if they have transferred the Ether or stable tokens to known scam addresses of the pool \textit{after} their removing liquidity transactions, we will flag them as collusion addresses.
    \item[3)] \textit{Swap valuable coins for the scam tokens.} 
    For the addresses that swapped Ether or stable coins for scam tokens (in order to raise the token price and attract victims), if they have ever received Ether or stable coins from known scam addresses of the pool \textit{before} their swapping transactions, we will flag them as collusion addresses. 
    \item[4)] \textit{Swap scam tokens for valuable coins.} For the address swapped scam tokens for Ether or stable coins to gain a profit, if they have transferred the valuable tokens to known scam addresses of the pool \textit{after} their swapping transactions, we will flag them as collusion addresses.
\end{itemize}

We believe these heuristics are comprehensive (i.e., covering most of the possible behaviors of collusion addresses on Uniswap) and reliable (i.e., by no means a victim would behave like this). 
Thus, from the known scam addresses (i.e., scam token/pool creators) of a given scam pool, we \textit{iteratively} discover the collusion addresses. At last, we get $41,118$ collusion addresses connected with $3,377$ scam token/pool creators. Among them, $39,758$ addresses had been used in swap operations, while $6,299$ addresses had been involved in liquidity related operations.

\subsubsection{Summary}
Overall, $70,331$ scam addresses related to the scams are identified, including $10,920$ scam tokens, $11,215$ scam pools, $6,288$ scam token creators, $6,465$ scam pool creators, and $41,118$ collusion addresses. Note that, one address can serve more than one roles across different scam liquidity pools.

\begin{table}[h]
\small
\centering
\caption{The top-10 profitable scam tokens and liquidity pools.}
\label{tab:top10token}
\resizebox{0.99\linewidth}{!}{
\begin{tabular}{@{}cccrr@{}}
\toprule
Pool Address                               & Token0                    & Token1               & \multicolumn{1}{c}{\begin{tabular}[c]{@{}c@{}}Profit\\ (\$)\end{tabular}} & \multicolumn{1}{c}{\begin{tabular}[c]{@{}c@{}}\# of \\ Victims\end{tabular}} \\ \midrule
0xfc2903fa0ee403b0e49cc7fb0919f04c4a49ee28 & certik.foundation (CTK)   & Wrapped Ether (WETH) & 1,502,118                                                                 & 365                                                                          \\ \midrule
0x0383eeb899e7fc0f4f696ebfcb5672ad7e0d271c & woo.network (WOO)         & Wrapped Ether (WETH) & 1,188,742                                                                 & 321                                                                          \\ \midrule
0xaa2e4317b13e3b4edfd45642516b31e211c3e71f & medicalveda.com (MVEDA)   & Wrapped Ether (WETH) & 871,450                                                                   & 142                                                                          \\ \midrule
0xa356939e22878af64560ba7e4253650f8cd9915d & flamingo.finance (FLM)    & Wrapped Ether (WETH) & 843,912                                                                   & 153                                                                          \\ \midrule
0xb7864c708ad58af75c756c26b1ba155bfa0e2307 & yfi.group (YFIG)          & Wrapped Ether (WETH) & 706,504                                                                   & 1,692                                                                        \\ \midrule
0xf2486c8f03afb444783427d620bf75510766e88d & akash.network (AKT)       & Wrapped Ether (WETH) & 628,409                                                                   & 119                                                                          \\ \midrule
0x57a5dd974adac8738d6796502c899d13e8903141 & Alpha Finance Lab (ALPHA) & Wrapped Ether (WETH) & 597,094                                                                   & 155                                                                          \\ \midrule
0x9e3fcc46ef41eb5c20f404c4c35848deb34044fc & Deriswap (DWAP)           & Wrapped Ether (WETH) & 498,349                                                                   & 124                                                                          \\ \midrule
0xaacd36c877408824ee59540b0c093804d7e9a7d9 & Meridian Network (MRDN)   & Wrapped Ether (WETH) & 489,992                                                                   & 923                                                                          \\ \midrule
0x700fa01ac5b01d6d92384062906f463292e682c9 & Injective Protocol (INJ)  & Wrapped Ether (WETH) & 477,553                                                                   & 135                                                                          \\ \bottomrule
\end{tabular}
}
\end{table}

\subsection{Measuring the Financial Impact}
\label{sec:impact} 
We next perform an impact analysis on these scams based on all the transactions related to Uniswap we collect in Section~\ref{sec:dataset_collection}. In each transaction, all the balance change related to participants on Uniswap are calculated and tracked. In total, these scam tokens profit over \$16 million, 
including over 28K ETHs and other leading official tokens, from $39,762$ potential victim addresses (i.e., all the scam liquidity pool investors excluding the scam addresses). 
On average, each scam liquidity pool has gained a profit of $\$1,477$. Table~\ref{tab:top10token} shows the top 10 most profitable pools. The most profitable scam liquidity pool, reserving \texttt{certik.foundation (CTK) - Wrapped Ether (WETH)} pair, made a profit of over \$1.5 million. It impersonated to be the official token of \texttt{Certik}~\cite{certik}, a blockchain security company whose official token contract was built on BSC chain, and fooled 365 potential victims. 
Since most of the scam tokens only have one liquidity pool (see Section~\ref{subsec:scampool}), the overall impact of scam tokens is similar to the liquidity pools, and the top-10 profitable tokens are also shown in Table~\ref{tab:top10token}. It is notable that all these 10 tokens are imitations of existing blockchain projects. 
Six of the top-10 scam tokens are camouflaging official ERC-20 tokens, while some scam campaigns also created counterfeit cryptocurrency targeting official tokens that were released on other blockchain platforms, such as \texttt{flamingo.finance} (FLM) on NEO blockchain.

\section{Discussion}
\subsection{Generality of Scam Tokens}

Although Uniswap V2 we study in this paper is one of the most representative DEXs based on its popularity, there are many other DEXs that operate well due to their unique features. Many DEXs, like Sushiswap, Bancor, etc., adopt the same/similar trading mechanism with Uniswap, i.e., the Automated Market Makers (AMMs). Thus, in practice, an attacker could easily create pools of scam tokens on multiple different DEXs to reach to as many unsuspecting investors as possible to gain more profit. 

\begin{table}[h]
\caption{The scam tokens of our dataset found in other DEXs.}
\centering
\small
\begin{tabular}{@{}crrc@{}}
\toprule
DEX        & \multicolumn{1}{c}{\# of Scam Tokens} & \multicolumn{1}{c}{\# of Total Tokens} & Launch Time \\ \midrule
Uniswap V3 & 5                                     & 3,183                                  & May-2021    \\
Uniswap V1 & 29                                    & 3,086                                  & Nov-2018    \\
Sushiswap  & 24                                    & 1,633                                  & Sep-2020    \\
Balancer   & 20                                    & 1,243                                  & Feb-2020    \\
Bancor     & 0                                     & 313                                    & Jan-2017    \\ \bottomrule
\end{tabular}
\label{tab:otherdex}
\end{table}

We provide an explorative study to show the generality of the issue and the prevalence of scam tokens on other DEXs.
We choose the other two versions of Uniswap and three popular DEXs (i.e., Sushiswap, Balancer, Bancor). We take advantage of the query APIs provided by these DEXs to investigate whether the scam tokens we identified have penetrated into DEXs beyond Uniswap V2.
Table~\ref{tab:otherdex} shows the overall result.
As expected, the scam tokens have slipped into other DEXs, i.e., all the five DEXs we explored have found scam tokens, although there are not many tokens listed on them. 
For example, the fake \texttt{Radar} (RADR) token\footnote{Token address:0x2933d48ab6833cfb56de5252277339f941f47cdb} was firstly listed on Uniswap V2 on 2020 July 4th. Within the same day, the attacker profited from the pool by Rug Pull and then the token was listed on Balancer the next day. 
Due to dataset limitation, we did not cover the detailed results of other DEXs in this paper.
Nevertheless, the detection methods proposed in this paper could be applied to these DEXs directly.

\subsection{Implication}

Our observations suggest the urgency to identify and avoid the scam tokens on DEXs.

\subsubsection{The system designers/operators.}
The root cause of this kind of scam is that DEXs do not maintain any rules for token listing, and Ethereum does not regulate the naming schemes of scam tokens. Thus, there is a strong need to design policies to regulate the token releasing on Ethereum and the liquidity pool listing on DEXs. However, this may contradict the goal of DEX, a fully decentralized marketplace. 
We believe a token reputation system is needed to decrease the impact of scam tokens. Techniques like the ones proposed in this paper could be used to flag suspicious tokens and further help develop methods to detect token contracts with backdoors. Instead of simply blocking the transactions on suspicious tokens, these detection results can be further embedded in the DEX front-end to warn users when they try to engage with the high-risk tokens.

\subsubsection{Investors}
Further, awareness should also be raised among investors. Rather than searching for tokens or pairs on Uniswap (as the scam token names are confusingly similar to the official ones), the investors should rely on trusted sources like CoinMarketCap or the official sites of DeFi projects to make sure they are trading with the official tokens. Also, before diving into the liquidity pool, investors should carefully check the transaction history of the pool, and pay special attention to the coin skyrocketing in price within a short time. Besides, as we find backdoors in some scam tokens' contracts, investors with experience in reading contracts could also look out for traps written in their contract codes. 

\subsubsection{DeFi Project Team}
At last, for the operation team of a DeFi project, they should be aware of the scam token abuse (which would hurt their reputation), and regularly post public announcements to remind their investors.

\subsection{Limitation}
Our study carries certain limitations. 

\textit{First, our scam token detection framework relies on some heuristics and manual efforts for verification}. While these heuristics proved effective, we acknowledge that they are too strict that the compiled scam token list may be incomplete. Indeed, our machine learning classifier flags much more suspicious tokens, while to the best of our knowledge, it is non-trivial to verify them and we could identify no better alternatives. Therefore, the characterization study in this paper provides the lower-bound results of the scam tokens on Uniswap. Nevertheless, we have curated by far the largest scam token dataset which will be shared with the research community. 

\textit{Second, our machine learning based detection relies on the transaction history of scam token, which may limit the usage scenario of our approach}. However, we argue that our approach can flag the scam tokens with partial or only a few transactions. For example, we have identified and verified over 1,800 scam tokens with less than 10 transactions using the machine learning based approach. It suggests that our machine learning based approach can act as a whistle blower that identifies scam tokens at their early stage (before they create a huge impact). 
Further, for the identified scam tokens, the guilt-by-association based expansion method could help us identify more scam tokens created by the same campaigns, even the newly emerging ones with no transactions. Thus, the machine learning based approach and the manual summarized heuristics can be used in the real-world usage scenario to identify scams at their early stage and eliminate the impact they caused. 

\textit{Third, although we have tried our best to understand the workflow of these scams and reveal the scam campaigns behind them, there might be more complex operation networks of the scams we did not touch} (e.g., we did not track how they launder the scammed money). Thus, it is quite possible that there are many scam addresses we did not observe.

\section{Related Work}
\subsection{Research on Cryptocurrency Exchanges}
Some researchers are focused on the security issues of centralized cryptocurrency exchanges~\cite{kim2018risk,mccorry2018preventing,chohan2018problems,feder2017impact,moore2018revisiting,ji2020deposafe}. For example, Kim et al.~\cite{kim2018risk} analyzed vulnerabilities of cryptocurrency exchanges and individual user wallets and Ji et al.~\cite{ji2020deposafe} demystified the fake deposit vulnerability related to exchanges and tokens.
Others are take efforts to evaluate or improve the effectiveness and reliability of decentralized cryptocurrency exchanges~\cite{lo2020uniswap,capponi2021adoption,wang2020automated,baum2021p2dex,annessi2021improving}. Lo et al.~\cite{lo2020uniswap} verify the effectiveness of decentralized exchanges and Annessi et al.~\cite{annessi2021improving} are exploring ways to improve security for DEX users through multiparty computation.

\subsection{DeFi Security}
The security of DeFi is also a hot research topic. There are many researchers studying on the price manipulation of DeFi~\cite{sobol2020frontrunning,boonpeam2021arbitrage,wang2021cyclic,tatabitovska2021mitigation,wu2021defiranger,qin2020attacking}. For example, Boonpeam et al.~\cite{boonpeam2021arbitrage} investigate arbitrage strategies and factors for profit-maximizing on decentralized exchanges, while Wang et al.~\cite{wang2021cyclic} focus on cyclic arbitrage on Uniswap and evaluate its impact.
Others have studied secure vulnerabilities and attacks on DeFi and Oracle~\cite{hsu2021defi,werner2021sok,gudgeon2020defi,oosthoek2021flash,caldarelli2021blockchain,wang2021blockeye}. For example, Hsu et al.~\cite{hsu2021defi} explored how design weaknesses in DeFi protocols could lead to a DeFi attack and 
Wang et al.~\cite{wang2021blockeye} proposed a real-time attack detection system for DeFi projects on the Ethereum blockchain through symbolic reasoning on smart contracts and monitoring transactions.

\subsection{Blockchain Scams}
Many kinds of blockchain scams have been studied, including the Ponzi Schemes~\cite{chen2018detecting, bartoletti2020dissecting, bartoletti2018data,vasek2018analyzing,chen2019exploiting,toyoda2019novel,bian2021scam}, fraudulent Initial Coin Offering (ICO)~\cite{liebau2019crypto,zetzsche2017ico}, phishing scams~\cite{wu2020phishers,phillips2020tracing,chen2020phishing}, bitcoin generator scams~\cite{badawi2020generator}, fake cryptocurrency exchanges~\cite{xia2020characterizing} and counterfeit tokens~\cite{gao2020tracking} , etc. Some of them also used machine learning methods to detect scams. For example, 
Badawi et al.~\cite{badawi2020generator} utilized search engines to search for web pages and train a classier to detect Bitcoin generator scams. 
Wu et al.~\cite{wu2020phishers} proposed a network embedding algorithm to identify phishing addresses. 
Despite this, as a kind of emerging scams, scam tokens on DEX have not been systematically studied yet and existing techniques cannot be applied to identify scam tokens directly.

\section{Conclusion}
This paper presents the first in-depth analysis of scam tokens on Uniswap. We have proposed an effective and accurate method for detecting scam tokens, and identified over 10K scam tokens and scam liquidity pools on Uniswap V2. We have systematically analyzed the scam behaviors, their working mechanism, and the financial impacts.
We reveal that scams are prevalent on Uniswap, and we speculate that similar scams could have been sneaked into other DEXs and Defi projects, because the inner cause lies in the loose/empty regulation of cryptocurrency on decentralized platforms. We advocate the cryptocurrency community to maintain a token reputation system using techniques like the ones proposed in this paper to eliminate the impact of scam tokens.

\section*{Acknowledgment}

We sincerely thank all the anonymous reviewers for their valuable suggestions and comments to improve this paper. This work was supported by the National Natural Science Foundation of China (grants No.62072046, 61772308, 61972224 and U1736209), Hong Kong RGC Project (No. PolyU15219319), the National Science Foundation under the grants CNS-2028748 and CCF-2046953. Haoyu Wang (haoyuwang@bupt.edu.cn) is the corresponding author.

\balance

\bibliographystyle{plain}
\bibliography{cite}

\appendix
\renewcommand{\thesection}{Appendix~\arabic{section}}
\setcounter{section}{0}

\section{Appendix for machine-learning based detection and verification in Section~\ref{sub:classifier}}
\subsection{Features used in our machine learning classifier.}
Table~\ref{tab:feature} shows the 40 kinds of features we extract to train a scam token classifier, including 7 kinds of time-series features, 24 kinds of transaction features, 4 kinds of investor features and 5 kinds of Uniswap specific features.

\begin{table*}[h]
\caption{The features used in our scam token classifier.}
\resizebox{0.99\textwidth}{!}{
\begin{tabular}{|c|cc|}
\hline
                              & Feature               & Description                                                                                         \\\hline
\multirow{7}{*}{Time-series}         & $T_{period}$         & The time interval from the first transaction to the last transaction on Uniswap      \\\cline{2-3}
                              & $T_{interval}$       & The time interval between the last transaction and the study time on Uniswap         \\\cline{2-3}
                              & $P_{mint}$       & The time point of mint events in the whole token lifecycle on Uniswap        \\\cline{2-3}
                              & $P_{swap}$        & The time point of swap events in the whole token lifecycle on Uniswap       \\\cline{2-3}
                                & $P_{swapfrom}$        & The time point of swap-from events (swap from tartget token for other token) in the whole token lifecycle on Uniswap      \\\cline{2-3}
                                  & $P_{swapto}$        & The time point of swap-to events (swap from other token for target token) in the whole token lifecycle on Uniswap       \\\cline{2-3}
                              & $P_{burn}$        & The time point of burn events in the whole token lifecycle on Uniswap     
                              \\\cline{1-3}
\multirow{24}{*}{Transaction}       & $N_{TxU}$              & Total transaction numbers on Uniswap                                                    \\\cline{2-3}
                              & $N_{TxE}$             & Total transaction numbers on Ethereum                                                   \\\cline{2-3}
                              & $N_{mint}$            & Total mint event numbers on Uniswap                                               \\\cline{2-3}
                              & $N_{swap}$            & Total swap event numbers on Uniswap                                               \\\cline{2-3}
                              & $N_{swapto}$          & Total swap-to event numbers on Uniswap                  \\\cline{2-3}
                              & $N_{swapfrom}$        & Total swap-from event numbers on Uniswap                     \\\cline{2-3}
                                                            & $RE_{swapfrom\_swapto}$        & $N_{swapfrom}/N_{swapto}$, set to -1 if the $N_{swapto}$ is 0                     \\\cline{2-3}
                              & $N_{burn}$            & Total burn event numbers on Uniswap                                               \\\cline{2-3}
                                                
                              & $A_{mint}$          & Total number of addresses that have participated in mint events on Uniswap        \\\cline{2-3}
                              & $A_{swap}$            & Total number of addresses that have participated in swap events on Uniswap        \\\cline{2-3}
                              & $A_{swapto}$          & Total number of addresses that have participated in swap-to events on Uniswap      \\\cline{2-3}
                              & $A_{swapfrom}$        & Total number of addresses that have participated in swap-from events on Uniswap    \\\cline{2-3}
                              & $A_{burn}$           & Total number of addresses that have participated in burn events on Uniswap        \\\cline{2-3}
                                 & $A_{all}$             & Total number of addresses that have participated in events on Uniswap                \\\cline{2-3}
                              & $RE_{mint\_all}$       & $N_{mint}/N_{TxU}$                                                                            \\\cline{2-3}
                           &  $RE_{swap\_all}$       & $N_{swap}/N_{TxU}$                                                                             \\\cline{2-3}
                        &  $RE_{swapto\_all}$       & $N_{swapto}/N_{TxU}$                                                                        \\\cline{2-3}
                               
                               &  $RE_{swapfrom\_all}$       & $N_{swapfrom}/N_{TxU}$                                                                       \\\cline{2-3}
                               &  $RE_{burn\_all}$       & $N_{burn}/N_{TxU}$                                                                       \\\cline{2-3}
                             &  $RA_{mint\_all}$       & $A_{mint}/A_{all}$                                                                   \\\cline{2-3}
                              &  $RA_{swap\_all}$       & $A_{swap}/A_{all}$                                                                   \\\cline{2-3}
                              &  $RA_{swapto\_all}$       & $A_{swapto}/A_{all}$                                                                     \\\cline{2-3}
                             &  $RA_{swapfrom\_all}$       & $A_{swapfrom}/A_{all}$                                                                  \\\cline{2-3}
                               &  $RA_{burn\_all}$       & $A_{burn}/A_{all}$                                                                     \\\cline{1-3}
\multirow{4}{*}{Investor} & $L_{mint/burn}$             & The average liquidity pools the participants that have minted or burnt on Uniswap                       \\\cline{2-3} &
 $L_{swap}$              & The average liquidity pools the participants that have swapped on Uniswap                       \\\cline{2-3}
                              & $C_{ming/burn}$         & The average mint or burn event counts of participants on Uniswap 
                              \\\cline{2-3}
                              & $C_{swap}$            & The average swap event counts of participants on Uniswap \\\cline{1-3}
\multirow{5}{*}{Uniswap Specific}        & $N_{pool}$                 & The number of liquidity pools                                                                      \\\cline{2-3}
                              & $V_{token}$         & Amount of tokens traded all time across pairs                                        \\\cline{2-3}
                              & $V_{tracked}$      & Amount of tokens in USD traded all time across pairs (only for tokens with a certain level of liquidity)                                            \\\cline{2-3}
                                & $V_{untracked}$      & Amount of tokens in USD traded all time across pairs (all tokens)                                              \\\cline{2-3}
                              & $N_{liquidity}$       & Total amount of token provided as liquidity across all pairs                                      \\\hline
\end{tabular}
}
\label{tab:feature}
\end{table*}

\end{document}